# Functional trends and rheological evaluation of polyurethane microcapsules in dermato-cosmetic applications

**Sharadwata Pan (শারদ্বত পান)[1,2,*,a], Andreas Wierschem[2], Natalie Germann[3], Thomas Becker[1]**

[1]Chair of Brewing and Beverage Technology, Technical University of Munich, Weihenstephaner Steig 20, D-85354 Freising, Germany
[2]Institute of Fluid Mechanics, Friedrich-Alexander-Universität Erlangen-Nürnberg (FAU), 91058 Erlangen, Germany
[3]University of Stuttgart, Institute of Process Systems Engineering, Böblinger Str. 78, 70199 Stuttgart, Germany

[a]Current address: Institute of Fluid Mechanics, Friedrich-Alexander-Universität Erlangen-Nürnberg (FAU), 91058 Erlangen, Germany.

## ABSTRACT

To date, natural and synthetic polymer-based microcapsules have been used extensively in various dermato-cosmetic applications, with an emphasis on the targeted delivery of active ingredients, including therapeutic and aesthetic interventions. Although numerous polymer candidates have been comprehensively investigated, polyurethane based microcapsules have received comparatively minor attention, despite possessing a multitude of intrinsic benefits. However, in recent years, although there has been an upsurge of studies involving polyurethane, predominantly as a capsule wall or shell component, towards tangible dermato-cosmetic applications, these are only intermittently documented. In the current review, we target this lacuna, explore, and collate only the most contemporary trends and advances (2017–to date) in the field. In addition, despite the significance, and pertaining to the acute deficiency of rheological studies targeting polyurethane-based microcapsules in dermato-cosmetic applications, we critically examine and lay a comprehensive interpretation, based on the current state-of-the-art, inevitability for systematic inquiries, and identification of several target domains that need urgent attention. Finally, we deliberate on the challenges and the impending projections from a diverse outlook. We focus on a steady and more sustainable path forward via incorporation of green raw materials, cumulative domain-optimized and customer-focused applications, and a significantly improved understanding of the microcapsule mechanical behavior via implementation of novel rheological characterization procedures.

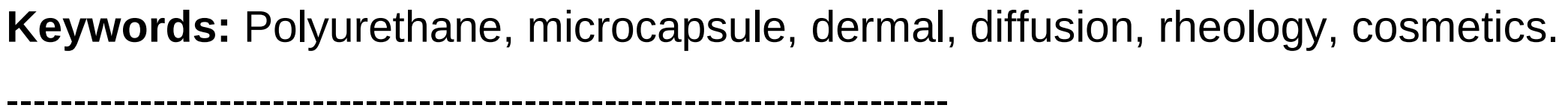

*Corresponding author.

Author E-Mails:
Sharadwata Pan (sharadwata.pan@fau.de),
Andreas Wierschem (andreas.wierschem@fau.de),
Natalie Germann (natalie.germann@svt.uni-stuttgart.de ; natalie.germann@degruyterbrill.com);
Thomas Becker (tb@tum.de)

## 1. INTRODUCTION

Skincare constitutes an extensive interdisciplinary field encompassing both fundamental research endeavors and large-scale industrial applications. The present study emphasizes two principal domains of skincare—dermal therapeutic interventions and cosmetic formulations—examined within the contexts of laboratory-scale investigations, pilot-scale optimization, and commercial-scale implementation with consumer acceptance considerations. Moreover, these conceptual frameworks exhibit a rational convergence with broader biomedical intervention strategies. In its contemporary form, the field of dermato-cosmetics represents a conjunction of three essential scopes: first, the pursuit of comprehensive integration of sustainably sourced, environmentally benign, and "green" components; second, the progressive design of dermato-cosmetic formulations as dual-function systems capable of effectively serving as carriers for bioactive agents; and third, the implementation of optimized, embedded delivery strategies that ensure controlled and sustained release of these active compounds to the target site [**1-6**; and references therein]. In 2020-21, the worldwide skincare market was assessed to be around USD ~98–100 billion, and by 2028, is projected to attain a staggering USD ~145 billion, with major market dominance shared by the Asia-Pacific (37%), North America (25%) and Europe (24%) [**7**]. This trend is also validated via comprehensive investigations carried out in recent years, from drug delivery to scaffold design, from smart textiles to shape memory biomaterials, from novel cosmeceuticals to therapeutic coatings, amongst others (see sections 2 and 3). Amongst numerous facets, a substantial one concerns the fabrication of innovative dermato-cosmetic materials along with the targeted delivery of active ingredients via strategies that would uphold their structural configuration, thermomechanical and physicochemical properties, while simultaneously ensuring a regulated release. To this end, microencapsulation of active agents and their subsequent delivery has garnered a perpetual research emphasis. Microencapsulation is the process of confining, enclosing or embedding an active agent or core inside an outer coating, matrix or shell material. The process offers tangible benefits to the encapsulated material, including protection against external stimuli, sustained release of the active agent or creating an active depot on the host target organ, thereby improving bioavailability. The resulting capsules or carrier vehicles are referred to as microcapsules, which vary in size, within a broad range of 1 µm to 1 mm in diameter. Particularly, there has been an upsurge in studies and reports documenting the benefits and shortcomings of polymer-based microcapsules, fabricated via synthetic and/or bio-based prepolymers or monomers, towards dermato-cosmetic applications [**8-14**; and references therein].

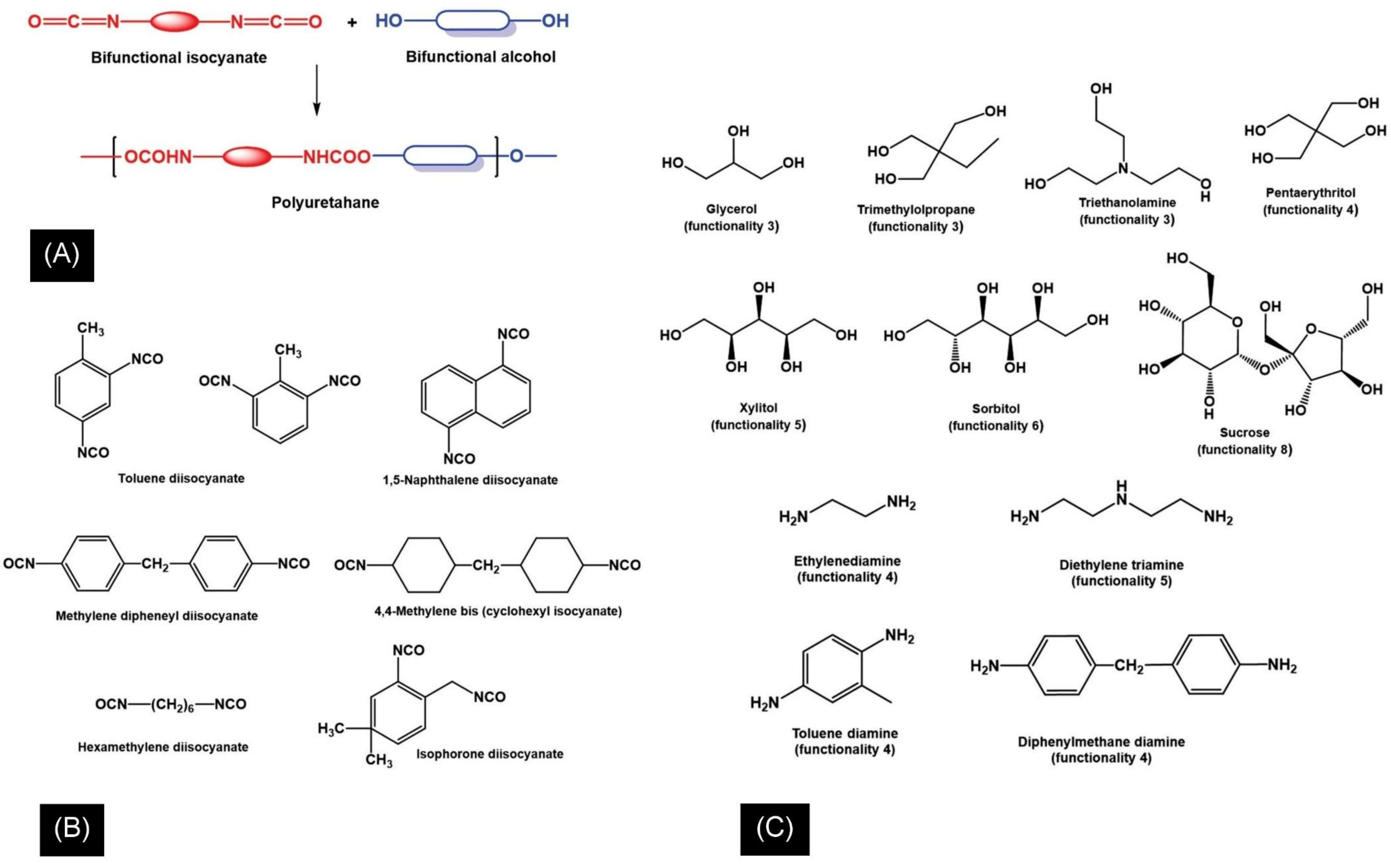

**FIGURE 1.** Chemistry of Polyurethane (PU). **(A):** Typical PU synthesis via an addition reaction between a polyol and a bifunctional isocyanate. **(B):** Representative isocyanates for PU synthesis. **(C):** Representative starter polyols for PU synthesis including their functionality. **(A)-(C):** Reprinted with permission from De Souza et al. [**15**]. Copyright (2021) American Chemical Society.

In the context of polymer-based microcapsules and their applications in the aforementioned sectors, polyurethane (PU) assumes an important role. This can be attributed to several of its intrinsic benefits: biocompatibility, ability to substitute synthetic with bio-based raw material (chain extenders, prepolymers, isocyanates, polyols) counterparts and their chemistry with the phase change materials (PCM) (**Fig. 1**), enhanced thermal stability, ability to confer a 'shape memory' influence, receptivity to external stimuli and/or environment, tunable surface properties via utilization and alteration of functional groups, core-monomer ratio and crosslinking agents, ability to generate diverse structural configurations (mono- or polynuclear conformation, spheres, core-shell structures: single, double or multiple-walls), propensity towards composite synthesis with novel nanomaterials (nanoparticles, nanotubes), amongst others [**16-26**; and references therein]. Despite the wide range of advantages, the utilization of PU based encapsulated delivery agents was scarcely reported compared to other polymers, both synthetic and bio-based. The usual PU microcapsule characterization procedures have effectively highlighted the application of numerous high-throughput methods, including optical techniques (scanning and transmission electron microscopy – SEM and TEM, atomic force microscopy – AFM), dynamic light scattering (DLS), thermal analysis via thermogravimetric analysis (TGA) and differential scanning calorimetry (DSC), tensile strength and texture characterization, chemical analysis via Fourier transform infra-red (FTIR) spectroscopy, gas and high-performance liquid chromatography (GC/HPLC), and mass spectrometry (MS), amongst others. These techniques offer relevant information suited to their end applications, including textile adhesiveness, principal component analysis, retention factors, encapsulation efficiency, transition enthalpy and specific heat capacity, wall thickness, surface charge, shape interpretation and size distribution. Conversely, insufficient mechanical characterization, poor correlation of the mechanical data with diffusion characteristics of the PCMs via *in vitro* release or penetration studies, and inadequate tunability of mechanical properties of the PU-based microcapsule suspensions to their desired applications, collectively highlight critical knowledge gaps that warrant systematic and comprehensive investigation.

From the purview of mechanical characterizations, in recent years, there is an increasing emphasis towards bulk rheological measurements [**27-33**]. Contextually, vesicle or capsule rheology could be foreseen analogous to cell rheological characterizations, mainly in terms of critical stress load limits, albeit demonstrating wide variations in their size distribution in suspensions [**34-37**]. As with other mechanical characterization protocols, rheology combines several characteristics simultaneously, for instance studying the influence of diverse, independent testing parameters using both stress- or strain regulated approaches, via different geometries and configurations, and/or implementing deformation testing under both steady and oscillatory shear as well as uniaxial and extensional modes. It additionally offers valuable insights onto the material behavior under small to large-scale deformations, simulating storage to processing conditions, either independently or in conjunction with microstructural interpretation via optical or rheo-optical techniques [**2**, **38-41**]. Despite the benefits, prevailing rheological assessments of PU-based microcapsule suspensions are extraordinarily limited, with minimal, standard investigations. Furthermore, tailored rheological characterizations towards end applications, processability and scale-up, although strictly necessary, are virtually non-existent (see section 4). Consequently, this also presents opportunities to lay a roadmap towards exhaustive rheological assessments of such microcapsules towards dermato-cosmetic applications.

In the current review, we collate the trends and outcomes from very recent studies, targeting the dermato-cosmetic applications of PU based materials, especially focusing on microcapsules (sections 2 and 3). In addition, we distinguish and document numerous inconsistencies, lacunae and challenges. Correspondingly, we present salient arguments and directions towards a more intense and miscellaneous rheology assisted strategy to interpret mechanical response of PU based microcapsules, which could eventually be suited to their end applications (section 4). Finally, we lay remarks on the forthcoming prospects from diverse perspectives (section 5) and summarize our key inferences (section 6).

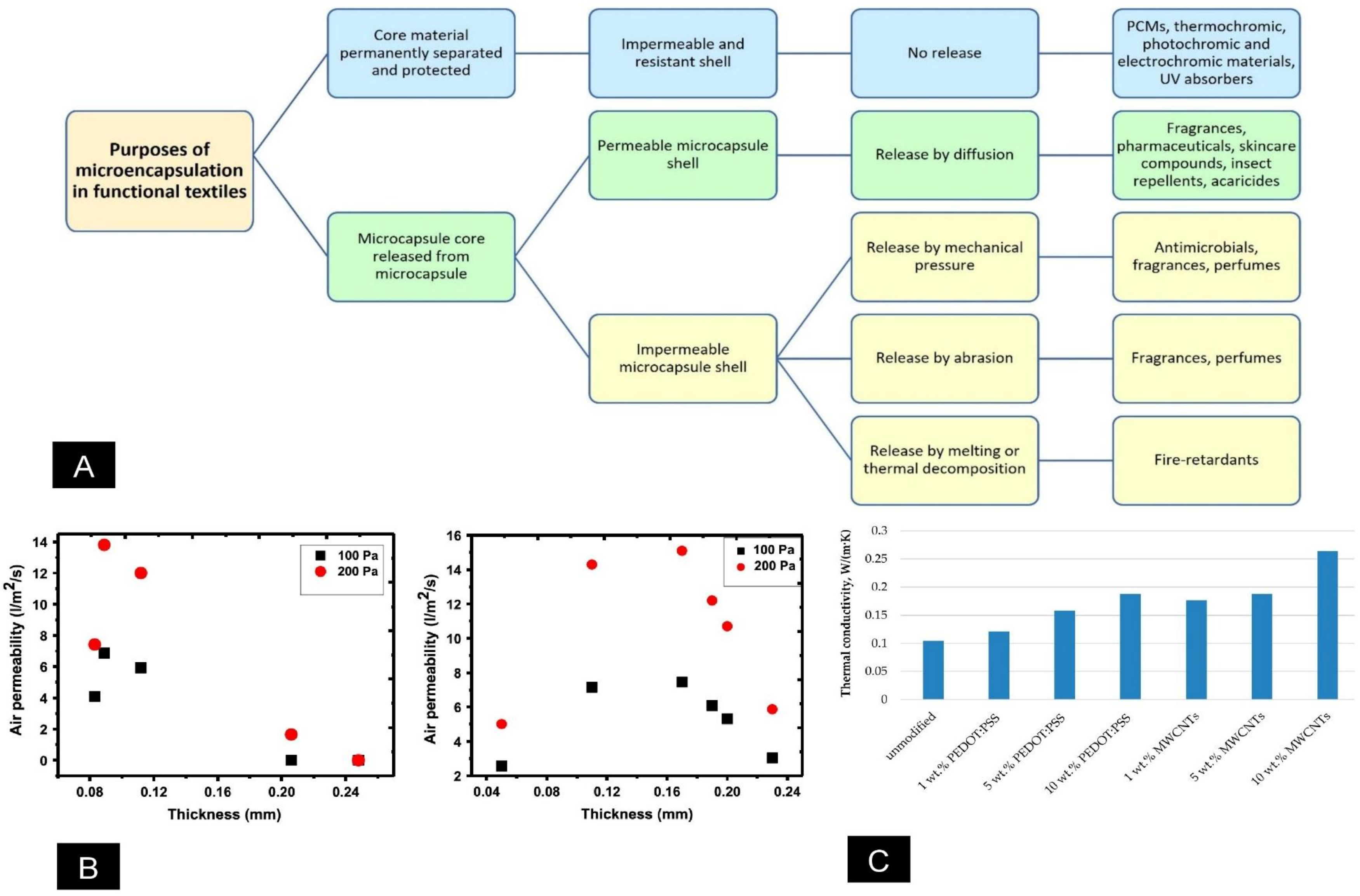

**FIGURE 2.** Textile applications of PU and PU based microcapsules. Part I. **(A):** Principal determinations towards the administration of PU/polymer-based microcapsules in functional textiles, from the purview of discharge machinery, PCM / bioactive constituents, and capsule wall porousness. Reprinted from Boh Podgornik et al. [**14**]. Open Access. Copyright (2021) The Authors. **(B):** Air permeability investigations. PU (left) and polyvinylidene fluoride (right) nanofibrous coatings obtained via electrospinning, embedded with silica aerogel. Reprinted with permission from Venkataraman et al. [**42**]. Copyright (2018) John Wiley & Sons, Ltd. **(C):** Heat conductivity of unchanged and wall-modified PCM (paraffin) microcapsules MPCM32D in a PU matrix. Here, PEDOT-PSS refers to poly (3,4-ethylenedioxyoxythiophene) poly (styrene sulphonate) and MWCNT refers to multiwall carbon nanotubes. Reprinted from Skurkyte-Papieviene et al. [**12**]. Open Access. Copyright (2021) The Authors.

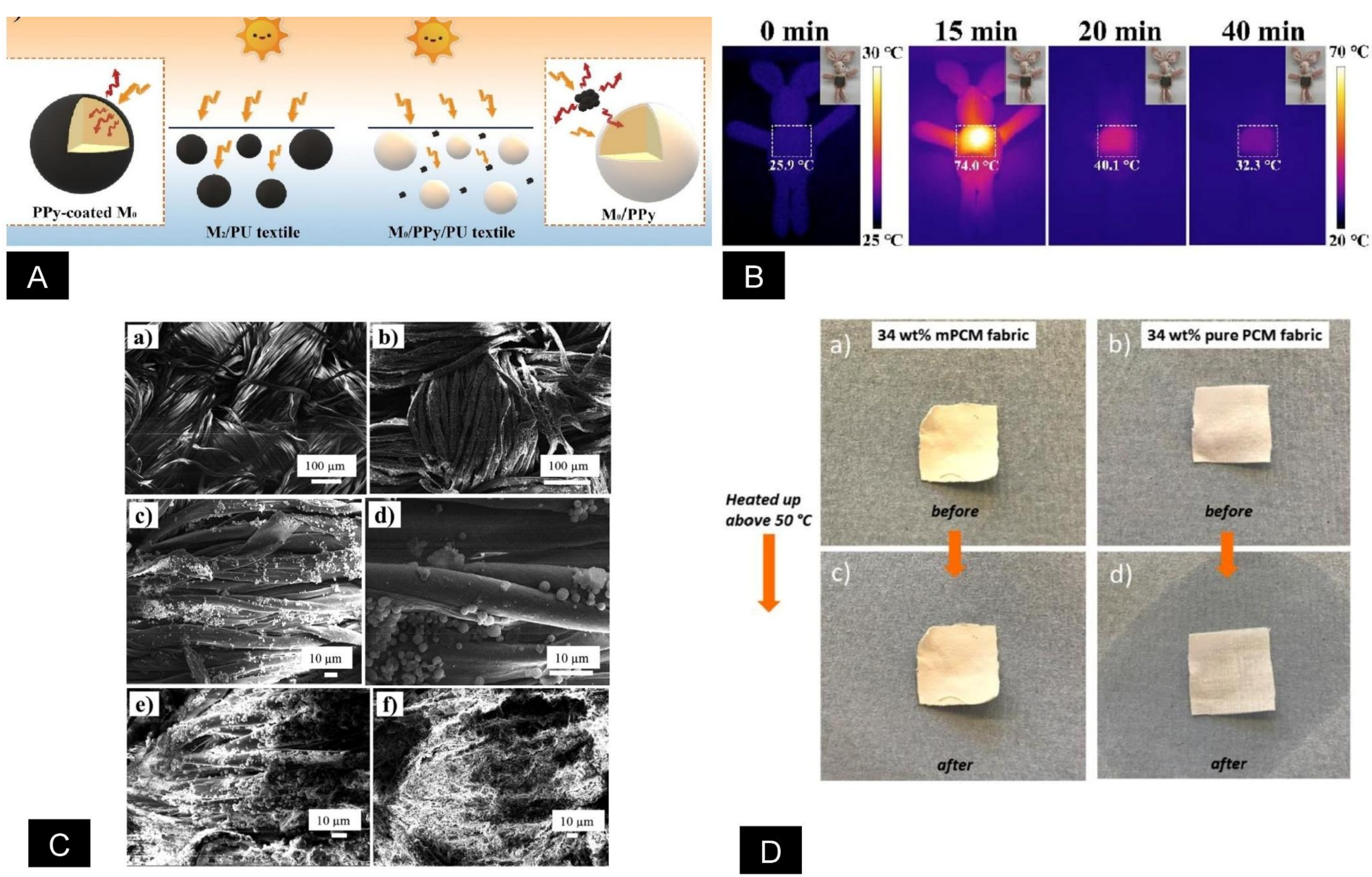

**FIGURE 3.** Textile applications of PU and PU based microcapsules. Part II. **(A):** Graphic display of the augmentation in photothermal proficiency of M2/PU textile vs M0/PPy/PU textile. Here PPy refers to polypyrrole (**B):** Illustration of heat therapy facilitated via PU based textile. **(A) and (B):** Reprinted with permission from Hu et al. [**43**]. Copyright (2021) Elsevier B.V. **(C):** SEM micrographs showing the cotton fabric (a) and various mass proportions of microencapsulated PCM or mPCM (PCM: n-docosane) administered into the cotton fabrics: (b−d) 2 wt.% (at diverse amplifications), (e) 8 wt.% and (f) 34 wt.% of mPCMs. **(D):** Photos of the cotton fabric with the highest mPCMs proportion, i.e., 4 wt % (a) prior to and (c) post being heated exceeding 50 °C, as well as of the fabrics with pure ndocosane (34 wt.%) (b) prior to and (d) post being heated to 50 °C. **(C) and (D):** Reprinted from De Castro et al. [**44**]. Open Access. Copyright (2021) American Chemical Society.

## 2. WEARABLE AND PERSONAL CARE COSMETICS

### 2.1. Cosmetotextiles

As an outline, interested readers may refer to the recent reviews [**11**, **45-49**]. Particularly, Petrulis and Petrulyte [**13**] have collated advancements of microcapsule applications towards the synthesis of novel fibrous materials, including an elaborate discussion on both fundamental structural configurations, as well as applications of PU microcapsules in cosmetotextiles. The review presents a diverse outlook via collation of a variety of synthesis and characterization techniques, and focusing on the application areas, which range from industrial processability, thermal conventions, revocable colour transformations, amongst others. Recently, Mertgenç et al. [**17**] synthesized PU urea (polyamine/PEG-400/isocyanate: varying molar ratios) based microcapsules towards textile applications, while encapsulating methyl cedryl ketone via interfacial polymerization. A key focus was to study the adhesion durability of the active ingredient on a range of fabrics, including polyester, non-woven fabric, cotton and silk. The authors reported a capsule wall size in the low micron range (~1μm), reasonable encapsulation efficiency (~70%), combined with a marginalized loss (~5%) of the active substance under ambient conditions after around two months and after 4 cycles of washing. Two observations were important: small sized PU microcapsules demonstrated a better heat stability based on TGA analysis, and the silk fabric showed the highest PU microcapsule retention rate. Incidentally, very recently, Xiao et al. [**11**] have reviewed the application of microcapsules in silk fibres, and collated data on both PU, as well as other biopolymers, namely, gelatin, chitosan, starch derivatives (such as $\beta$-cyclodextrins), as coating polymeric materials in textile finishing. The authors noted that while PU based microcapsules present benefits in the form of elevated tensile strength, tissue compatibility, microbial resistance, they also bear shortcomings in being too breakable and show poor heat resistance.

In addition, the role of polymeric materials in synthesizing potent, novel nanocomposites, for instance PU and silver-nanoparticles, towards a significantly better adhesion between the silk fibres and the microcapsule walls, mainly via either chemical bonding or interactions between molecules, has been emphasised. The resolutions and suitability of PU microcapsules in textile finishing have also been highlighted (**Fig. 2A**) recently by Boh Podgornik et al. [**14**], who stressed on both the chemical and physical manipulation and the synthesis strategies, including solvent evaporation, interfacial and *in situ* polymerization, simple and complex coacervation techniques. Importantly, the study focused on environmental sequestration aspects, arising from the recent attempts of administering naturally occurring biopolymers towards PU microcapsule synthesis strategies towards textile applications. The authors report that although recent years have seen increasing progress upholding this trend, persistent challenges in the form of prolonged solidity and coating robustness create obstacles that warrant attention. Electrospinning is also a viable strategy to synthesize innovative nanofiber membranes towards textile applications focusing on thermal insulation, as shown by Venkataraman et al. [**42**], via the manufacturing of malleable polyvinylidene fluoride and PU nanofibers implanted with silica aerogel (**Fig. 2B**). The authors validated that although PU showed lesser thermostability in relation to polyvinylidene fluoride, in general, heat resistance could be advanced by increasing the duration of the electrospinning procedure, or/and via enhancement in both the mass/area and the number of nanofiber membranes. Skurkyte-Papieviene et al. [**12**] have recently demonstrated that the incorporation of PU and multiwall carbon nanotubes in the microcapsule shells as excipients (optimal concentration: ~5wt.%), with paraffin as the PCM, increases heat conductivity by ~12% in textile applications, compared to the unaltered shell paraffin microcapsules, via usual thermal parameter-based analysis (**Fig. 2C**). Bio-based PU microcapsules have been reported recently by Azizi et al. [**21**], using $\beta$-cyclodextrin (rather than traditional diols) dependent PU shell and 2-ethoxynaphtalene or neroline (an active cosmetic ingredient) as the PCM via interfacial polycondensation method. The authors reported ~40% encapsulation efficiency, largely without aggregation, a monodisperse size distribution, and ~74% yield via saturation on polyamide interwoven fabric. The most interesting aspect is that the PU microcapsules were shown to retain their efficacy even after 35 washing cycles, indicating successful utilization of bio-based PU microcapsules towards textile applications. Earlier, Ben Abdelkader et al. [**50**] has also demonstrated the suitability of bio-based PU microcapsules for the delivery of neroline fragrance (**Figs. 4C and 5A**). The authors used hexane diisocyanate and isosorbide as starting monomers and fabricated the PU wall via interfacial polymerization. The resulting microcapsules, although showed a broad size range (2–100 μm), nevertheless demonstrated significant thermal tolerance (up to 260 °C), and a reasonable encapsulation efficiency (~73%). Even after 40 successive washing cycles, adequate proportion of residuary neroline was detected in the test cotton materials. Very recently, the same group reported $\beta$-cyclodextrin employed bio-based PU microcapsules for the delivery of neroline [**16**]. The study reported optimum impregnation circumstances for producing a unique case of weight accumulation (8.56%) in terms of surfactant (0.72 mg·mL$^{-1}$), cross-linker (31 mg·mL$^{-1}$) and microcapsule (21 mg·mL$^{-1}$) concentrations. Additional textile applications of PU and PU based microcapsules are shown in **Figs. 2** and **3**.

Inclusion of PU microcapsules into functional textiles may be achieved via diverse strategies [**14, 47, 51-53,** and references therein], including impregnation, stamping, immersion and padding, spraying, coating, amongst others. Of particular interest are the recent advances on spraying and immersion strategies **[47, 52]**, both of which necessitates a finishing protocol and the employment of a binder, which holds the microcapsules onto the textile or fabric even after utilization and washing. These strategies mainly deal with the PU microcapsule suspension, the chemical binder and excipients, such as softeners, either being sprayed on the textile surface or used as a submerging medium for the fabric. PU may be employed both as a capsule ingredient, or as the binder. For instance, Yılmaz and Karavana **[54]** reported spraying of chitosan microcapsules with 25-part aliphatic and non-ionic PU as a binder to fabricate eco-friendly leather for wearable footwear applications. On the other hand, fabrication of functional textiles via immersion and padding techniques of PU as a microcapsule shell material in varying concentrations has been reported towards different textile materials, including silk, polyester and silk **[17, 55]**. These studies demonstrate that while the application methodology varies, both spraying and immersing-padding strategies involve post processing via thermal fixation, and are influenced by the origin and type of PU as the binder, compatibility between the PU and the fabric, nature of PU microcapsule ingredients, degree of thermal regulation, amongst others. Comparative benefits and drawbacks are also highlighted. For instance, spraying PU capsules facilitates decorative applications for subtle or finished textiles, which does not require a smooth immersion. It is additionally essential to stabilize the cargo via heat (130 to 170 °C). However, upholding the diffusion height and consistency can be problematic. Here, immersion is preferable, which may result in a higher PU microcapsule application efficiency compared to spraying. Regardless of the procedure, a common concern is that fast swelling and evaporation may lead to destabilization of the capsule due to the breakage of the shell.

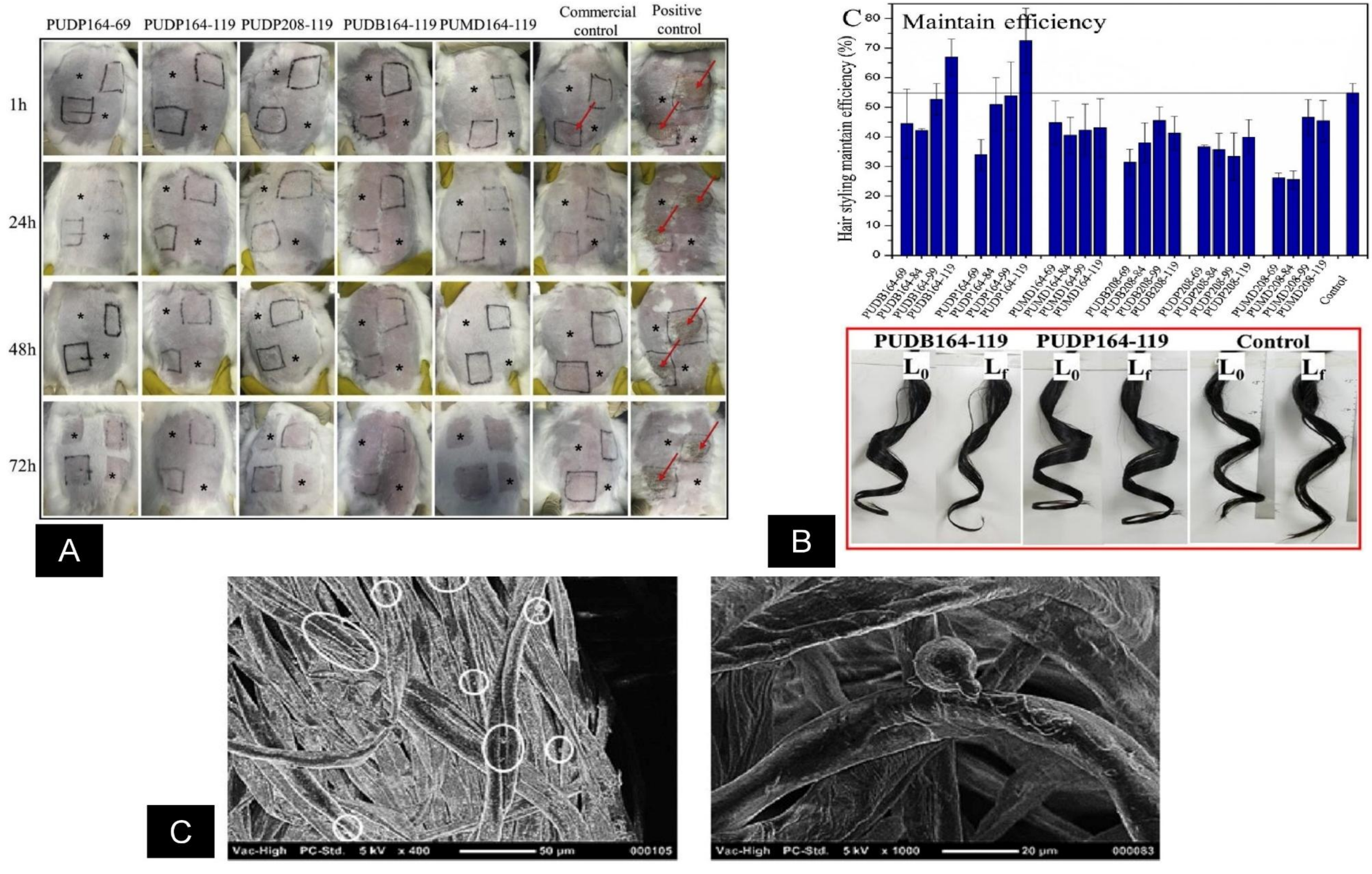

**FIGURE 4.** Cosmetic and cosmetotextile applications of PU and PU based microcapsules. Part I. **(A):** Dermal irritation examinations in rabbits following treatments with bio-based waterborne PU emulsions (with different ratios of the isocyanate and the chain extender). Macroscopic manifestation on the skin post elimination of the assessment bandages at different time intervals (hours), including the control specimens. The symbol (*) denotes the location of blank control. The red arrows designate swelling and redness. **(B):** (top) Hair-fashioning competence ratio and (bottom) specimens with robust hair fashioning preservation functioning post treatment with 3 wt.% bio-based waterborne PU emulsions, combined with hair spray, first at 120 °C, followed by equilibration at a raised humidity level (75±5% RH) for 1.5 hours. **(A) and (B):** Reprinted with permission from Zhang et al. [**56**]. Copyright (2020) Elsevier B.V. **(C):** Cosmetotextile applications of neroline-loaded bio-based PU microcapsules. SEM images of fragrant cotton-based functional fabric with microcapsules at two different magnifications. Reprinted with permission from Ben Abdelkader et al. [**50**]. Copyright (2018) Elsevier B.V.

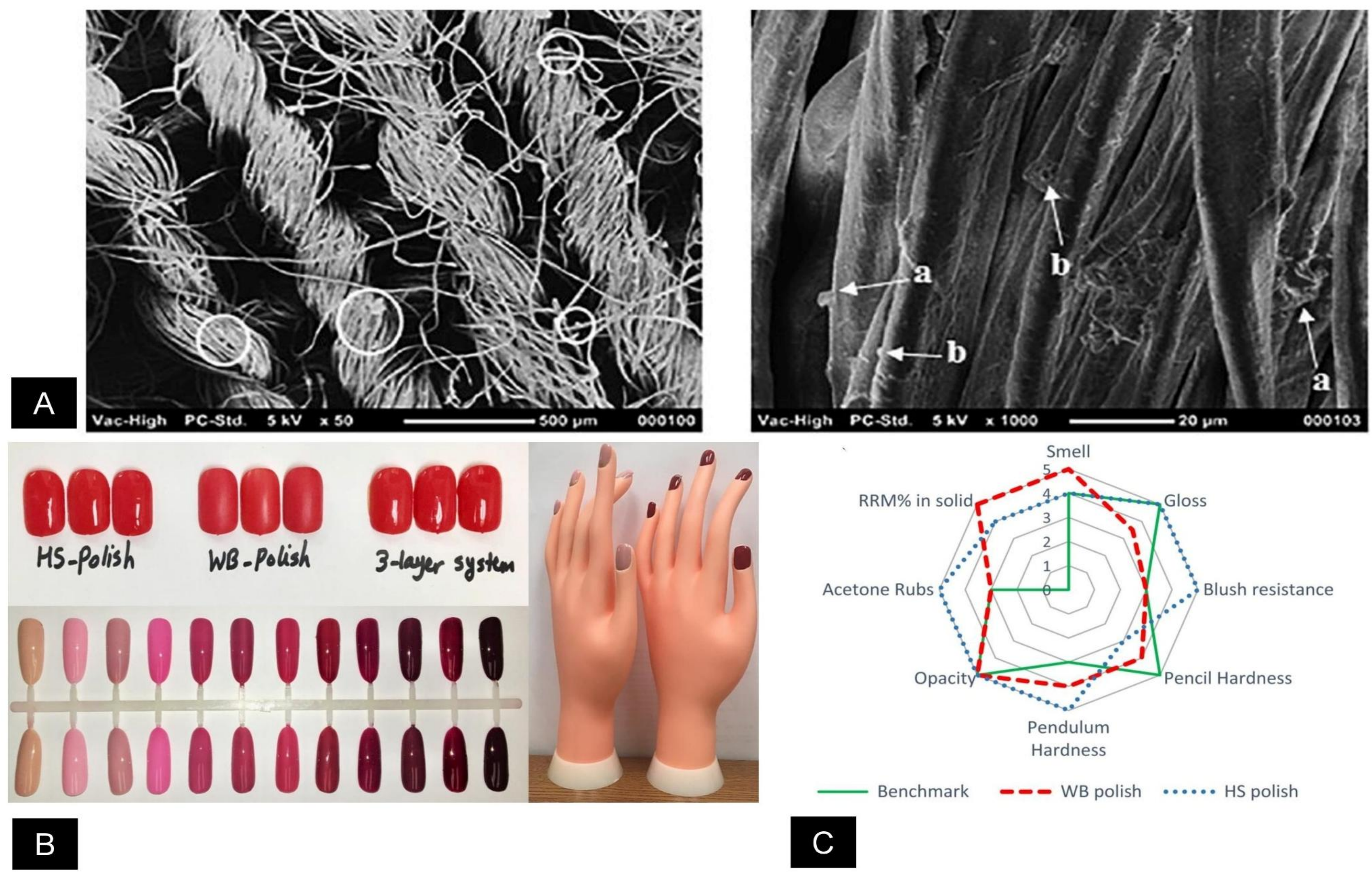

**FIGURE 5.** Cosmetic and cosmetotextile applications of PU and PU based microcapsules. Part II. **(A):** SEM images of the same, as in FIGURE 4(C), post 40 washing cycles at two different magnifications. Reprinted with permission from Ben Abdelkader et al. [50]. Copyright (2018) Elsevier B.V. **(B):** Manifestation of the gel nail polishes smeared on synthetic nails. Here, the WB gel nail polish contains bio-based, acrylated, waterborne PU dispersion. **(C):** Demonstration of significant technical advantages of the WB nail polish via the assessment of the inclusive functioning of different nail polishes, including the benchmark products. **(B) and (C):** Reprinted with permission from Zareanshahraki and Mannari [**57**]. Copyright (2018) Society of Cosmetic Scientists and the Société Française de Cosmétologie.

### 2.2. Personal care applications

Utilization of PU microcapsules has been primarily concentrated on fabric engineering and textile finishing, where their robustness and tunable permeability have proven advantageous. More recently, research within the personal care sector has expanded toward integrating PU microcapsules into advanced cosmetic formulations. In this context, PU microcapsules are being engineered to facilitate the controlled and targeted delivery of active ingredients—such as emollients, vitamins, and other functional biomolecules—directly to the skin surface. This emerging approach not only enhances ingredient stability and prolongs release profiles, but also enables improved sensory performance and efficacy, marking a significant evolution in the design of smart cosmetic delivery systems. For an overview, interested readers may refer to the recent reviews [**46, 58-65**]. Lu et al. [**26**] reported the synthesis of single-core, double-shell PU based microcapsules via interfacial polymerization for the delivery of PCMs, i.e., n-octadecane and butyl stearate (in an earlier study), two widely adopted active cosmetic ingredients in the personal care industry. Here, the outer and inner shells were made of polyurea and PU, respectively, synthesized employing mostly synthetic ingredients as monomers and additional starting materials. The reported microcapsules were in the lower size range threshold (~3-5 μm), with high encapsulation efficiency (>90%), compressed and smooth surface characteristics, and significant thermal stability. The latter was interesting, since PU based microcapsules, especially with large diameters and single shells, were usually perceived to possess inefficient heat tolerability. The efficacy of this strategy was also demonstrated later by Yin et al. [**25**], who have used the identical PCM, i.e., n-octadecane, and the synthesis procedure, i.e., interfacial polycondensation, but with minor changes in the choice of monomers. One important variation compared to the study by Lu et al. [**26**] was the incorporation of polyethylene glycol 400, which substituted polypropylene glycol (PEG) 2000. The authors reported significant morphological transformation for the microcapsules via SEM analysis, along with enhanced heat stability via DSC and TGA analyses, similar to the study by Lu et al. [**26**], as well as the use of PEG 400 to fabricate the PU shell polyurethane shell and adjust the morphology of the capsules. SEM images proved that with the modification of PEG 400, the microcapsules morphology improved significantly. Both studies by Lu et al. [**26**] and Yin et al. [**25**] clearly showed that multi-walled microcapsules with PU as one of the shell components could be an effective strategy to deliver cosmetic ingredients on a long-lasting basis. Additionally, although the recent study by Mertgenç et al. [**17**] showed suitability of PU based microcapsules towards efficient textile fabric finishing, as discussed earlier, this could also be an efficient way to deliver the key PCM, i.e., Methyl cedryl ketone, as an active fragrance ingredient in both aesthetic, as well as therapeutic cosmetic applications. Recently, there has been much focus on waterborne PU dispersions, i.e., aerogels, nanosuspensions, amongst others, based on both synthetic and bio-based ingredients as raw materials, and their applications in the area of personal care cosmetics, including sunscreen, nailcare and hairstyling products [**66**; and references therein]. In line with the existing trend, bio-based aqueous PU based dispersions have garnered attention as viable alternatives to their synthetic counterparts [**24**; and references therein]. A few cosmetic and cosmetotextiles applications of PU microcapsules are displayed in **Figs. 4 and 5**.

## 3. BIOMEDICAL APPLICATIONS

### 3.1. PU based materials

Utilizations of PU in various therapeutic applications, both of synthetic and biological origins, have been adequately documented. Interested readers may refer to the very recent reviews for an overview [**67-73**]. The detailed reviews by Wendels and Avérous [**74**] and Sikdar et al. [**75**] offer a comprehensive discussion on bio-based and synthetic PU based biomaterials and their applications. Several intrinsic advantages of PU are highlighted by both studies, including desirable mechanical characteristics and adaptability, straightforward operational flexibility, flexural durability and scratch endurance, which make it a versatile and desirable candidate towards potent and innovative therapeutic applications (subfigures in **Figs. 6 and 7**). This is also demonstrated via several additional recent studies, which have either documented or carried out experimental studies focusing on a wide range of applications of PU based micro- and nanomaterials in different forms, i.e., films, scaffolds, membranes, aqueous dispersions, foams, aerogels, amongst others, in different relevant areas, including tissue engineering and regenerative medicine, medical implant devices in invasive and non-invasive prognosis and care, scaffolds and wound healing, targeted drug delivery, medical consumables (such as disposable items, supplies), as well as other technical apparatuses. For instance, Feldman **[76]** provided a focused assessment of PU and PU-based nanocomposites in biomedical contexts, emphasizing that, beyond their well-established advantages, they are inherently bioinert. This is a fundamental attribute governing their biocompatibility and functional performance. Gupta et al. **[77]** highlighted the extreme versatility of shape-memory PUs and PU-derived composites, showcasing their remarkable self-healing capacity and mechanical robustness across diverse processing conditions. Building upon these advancements, Balavigneswaran and Muthuvijayan **[78]** demonstrated that innovative PU-based injectable and self-healing hydrogels exhibit superior stress tolerance and tunable mechanical integrity for tissue engineering applications, achieving up to a twentyfold enhancement in compressive strength. Similarly, Anıl et al. **[66]** surveyed the potential of waterborne PUs and nanoparticle-enriched PU dispersions for a broad spectrum of targeted biomedical applications, with particular emphasis on antibacterial coatings and implantable devices. Ma et al. **[10]** reviewed emerging phase change materials (PCMs) and polymer-based encapsulants across multiple technological domains, underscoring their relevance in next-generation biomedical and energy systems. Amongst others, the authors highlighted the effectiveness of a paraffin silica aerogel with PU as the outermost stratum towards designing innovative cold storage boxes for vaccine transportation, a PU based foam in medical dressing, as well as a PU/paraffin-based nanocomposite towards chemotherapy during carcinogenesis. Mohamady Hussein et al. [**79**] very recently reported the efficacy of PVA-Gelatin/PU nanofibrous scaffolds (containing nanoceria and essential oils from cinnamon) via electrospinning technique towards the treatment of diabetic wounds. The authors noted that the effectiveness of the novel nanofibrous scaffolds was pronounced more towards gram-positive bacteria (*Staphylococcus aureus*) rather than the gram-negative bacteria (*Escherichia coli*). A handful of applications of PU based (bio) materials are displayed in **Figs. 6 and 7**.

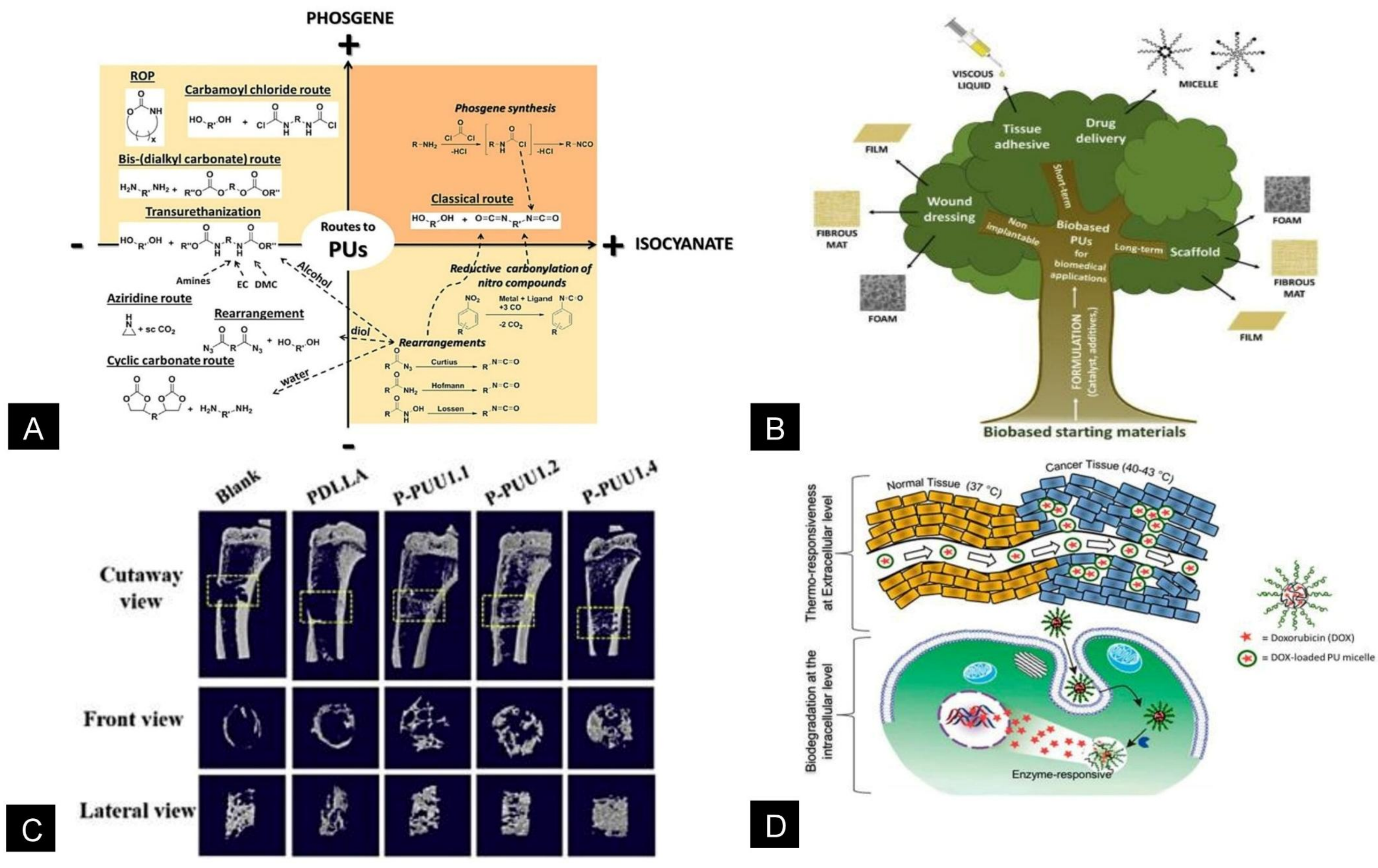

**FIGURE 6.** Biomedical applications of PU based materials. Part I. **(A):** Principal fabrication strategies for synthetic PU. **(B):** Fundamental biomedical applications of bio-based PU, including their predominant configurations. **(C):** Original bone generation via Micro Computed Tomography analysis following implantation (after 56 days): 3D restoration prototype of novel bone synthesis. P-PUU1.1–P-PUU1.3: PUs with accumulative proportions of piperazine. Reprinted with permission from Ma et al. [**80**]. Copyright (2019) American Chemical Society. **(D):** Drug discharge via PU based micelles towards anticancer therapy: dual-responsiveness of PU and general conception. **(A), (B) and (D):** Reprinted from Wendels and Avérous [**74**]. Open Access. Copyright (2020) The Authors.

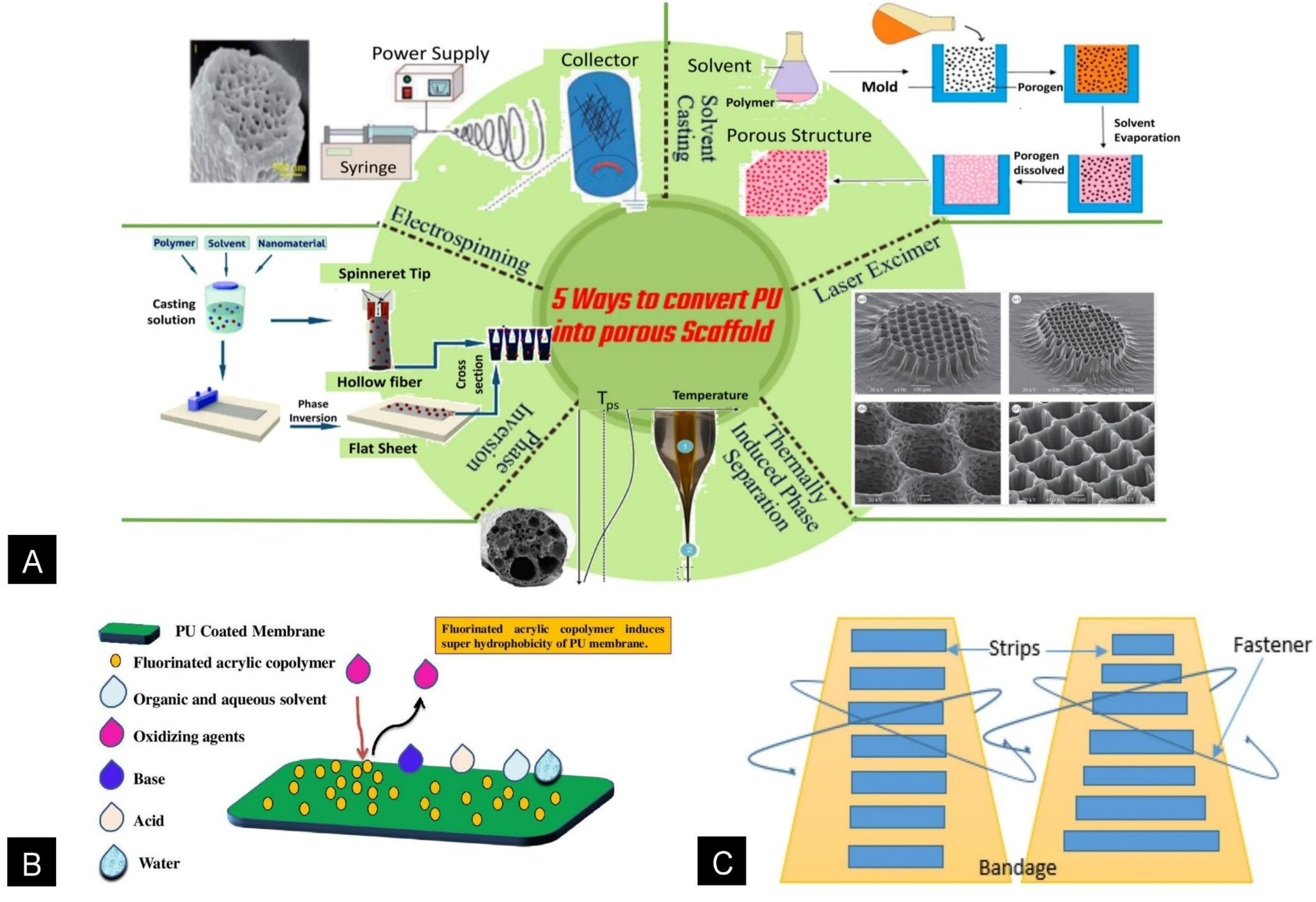

**FIGURE 7.** Biomedical applications of PU based materials. Part II. **(A):** Several distinct and diverse strategies to transform PU into permeable scaffolds. **(B):** Membrane smeared with PU with acrylic copolymer (fluorinated). **(C):** Shape memory PU based strips incorporated into pressure bandages. **(A)–(C):** Reprinted from Sikdar et al. [**75**]. Open Access. Copyright (2022) Wiley Periodicals LLC.

### 3.2. PU based microencapsulated PCMs

Despite these significant technical advancements concerning PU-based materials, incorporation of both bio-based and synthetic PU in the microencapsulated form is not as widely characterized and reported, compared to other delivery modes. Certainly, recent studies focusing on PU-based microcapsules towards dermato-cosmetic applications are mostly dominated by the cosmeceuticals, as discussed in the previous section. Nonetheless, a handful of recent studies do report on the proposed research direction [**53, 59, 81, 82**; and references therein). Selected applications of PU based microcapsules are presented in **Figs. 8 and 9**. Zhang et al. [**19**] designed novel configuration retention core-shell type PU microcapsules towards drug delivery applications via interfacial polycondensation approach, using a mixture of synthetic and bio-based monomers, emulsifier and starting materials (**Fig. 8A**). The authors reported that their strategy is scalable with significant upside potential in drug delivery. Additionally, the positioning of suitable chemical cross-linkers, such as one with triple functionalities, could principally control the shape memory effect, while increasing the recovery frequency of the desired microcapsules. Very recently, the suitability of PU based microcapsules as effective medium, short and sustained drug delivery agents targeting carcinogenic growths, was extensively reviewed [**22**; and references therein], which is arguably one of the most potent application areas (**Figs. 8B and 8C**). A significant impetus could be that majority of drug carriers, when administered in micro- or nano-encapsulated forms, positively influence the drug load and persistent release, related to other forms. Additionally, PU promotes cellular attachment and proliferation, with minimal or no side effects. For instance, earlier, Niu et al. [**83**] have reported multifunctional PU based microcapsules towards the delivery of doxorubicin, where cellular incorporation could be tracked, and the drug release could be regulated via changes in pH. Additionally, a large number of past studies have presented evidences in favour of PU in targeted drug delivery, although strictly not in microencapsulated forms, such as micelles, nanogels, nanoparticle-based composites, amongst others [**22**; and references therein].

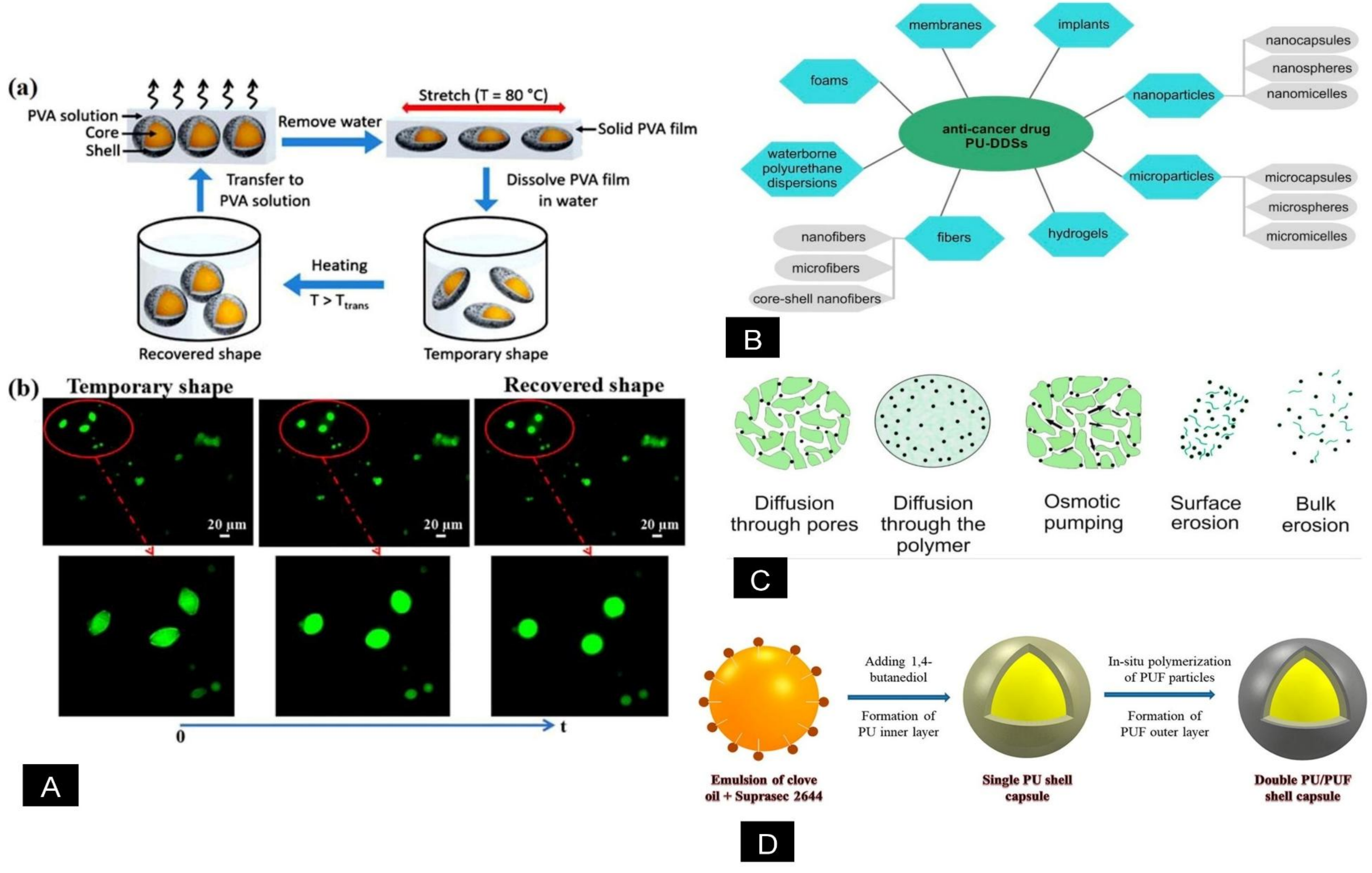

**FIGURE 8.** Biomedical applications of PU based microcapsules. Part I. **(A):** (a) Representative depiction of the retrieval and programming method of shape memory PU (SMPU). Initially, PU based microcapsules were incorporated in a polyvinyl-alcohol (PVA) film to facilitate a heat-stimulated stretching. The remote automated microcapsules are acquired post water dissolving of PVA. The microcapsule initial configuration is restored at temperatures exceeding the melting temperature. Consequently, the entire procedure can be reiterated. (b) Demonstrative imageries of the automated and shape retrieval performance (at 60 °C, duration: 30 s) of the SMPU microcapsules, as noted from the transformation of the aspect ratio (1.66: provisional, 1.06: regained). Reprinted with permission from Zhang et al. [**19**]. Copyright (2020) American Chemical Society. **(B):** Various PU based drug delivery schemes aimed towards targeted anti-cancer treatments. **(C):** Diverse drug discharge machineries concerned in dissimilar PU based / polymeric drug delivery schemes. **(B) and (C):** Reprinted from Sobczak and Kędra [**22**]. Open Access. Copyright (2022) by the authors. **(D):** Representative synthesis strategy of core-shell type PU microcapsules (double-shelled: PU/poly(urea-formaldehyde) or PUF; PCM: clove oil). Reprinted with permission from Chong et al. [**20**]. Copyright (2018) American Chemical Society.

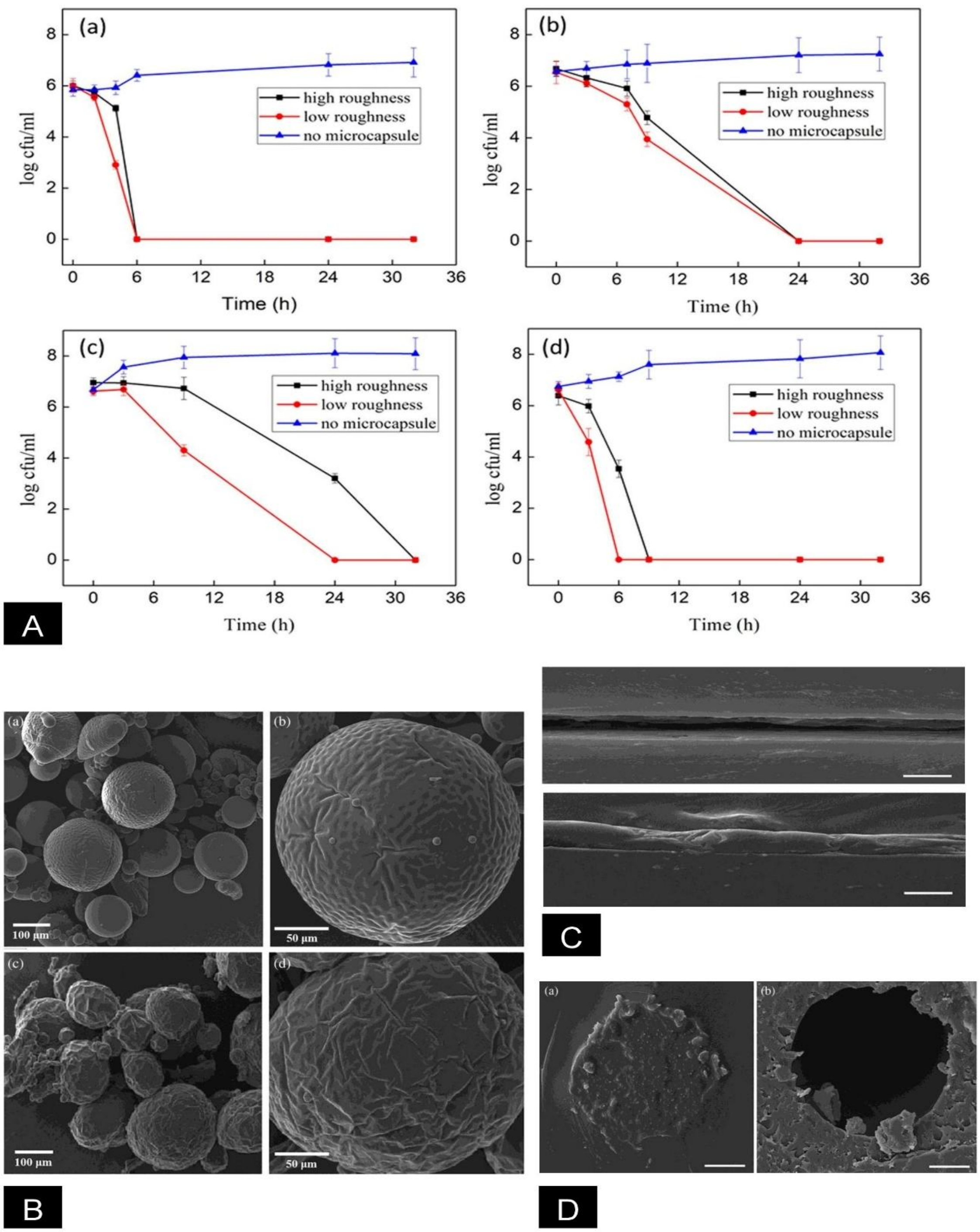

**FIGURE 9.** Biomedical applications of PU based microcapsules. Part II. **(A):** Antimicrobial actions of the same double-walled PU based microcapsules, as in FIGURE 8(D), against diverse bacterial species: *V. coralliilyticus* (a), *E. aestuarii* (b), *E. coli* (c), and oceanic biofilm-generating bacteria (d), derived from adulterated specimens on the spot. Reprinted with permission from Chong et al. [**20**]. Copyright (2018) American Chemical Society. **(B):** SEM images of M30 (c and d) and M10 (a and b) PU based microcapsules, with 30 wt.% and 10 wt.% epoxy with n-butyl acetate solvent, respectively. **(C):** SEM images of the scrabbled zone of epoxy coating specimens incorporated with (top) 0 wt.% (only epoxy coating) and (bottom) 20 wt% PU based M10 microcapsules. Scale bar represents 0.025 mm. **(D):** SEM images displaying microcapsule compatibility with (a) epoxy, and (b) PU layers (b) and epoxy (a). Scale bar represents 0.05 mm. **(B)**–**(D):** Reprinted with permission from Parsaee et al. [**23**]. Copyright (2020) Wiley Periodicals LLC.

Some or most of these cases could be investigated for potential scale up and full-scale utilization as microencapsulated vehicles. Chong et al. [**20**] synthesized double-shelled poly(urea-formaldehyde)/PU microcapsules clove oil as the PCM/core in oil-in-water emulsion via interfacial and *in situ* polymerization reactions. The authors noted that the microcapsules could find potential applications as antifouling agents, with targeted inhibitory actions against a wide range of bacteria (**Figs. 8D and 9A**), including marine biofilm forming bacteria, *E. coli*, and that the PCM release frequency could be regulated via the alteration of the amount of PU reactants. Parsaee et al. [**23**] reported the fabrication of epoxy resin containing globular (wall width: 0.6–0.7 μm; diameter: 50–720 μm), self-healing, core-shell type urea-formaldehyde/PU microcapsules with appreciable thermal stability via *in situ* polymerization reaction with targeted applications towards anti-corrosion coatings (**Figs. 9B**–**9D**). The authors noted that the corrosion resisting features of the coating layer could be enhanced just by enhancing the microcapsule load, and that even at 10% microcapsule load, the anti-corrosion features could be maintained for up to 35 days.

## 4. RHEOLOGICAL ASSESSMENTS

### 4.1. Current state-of-the-art

Widespread efforts have been devoted to the rheological characterization of PU based systems—including coatings, films, composites, and hydrogels—synthesized from both bio-based and synthetic constituents **[29, 30, 84-87]**. These studies reveal two dominant and complementary approaches to understanding their mechanical behavior. The first seeks to understand the behavior of PU materials under controlled unidirectional shear deformation, thereby obtaining fundamental steady-state parameters, such as shear viscosity and shear stress. The second focuses on the dynamic characterization of viscoelasticity via oscillatory shear experiments, wherein sinusoidal strain or frequency inputs yield parameters such as the storage modulus, loss modulus, and complex viscosity. Together, these methodologies offer a comprehensive framework for probing the structure–property relationships governing PU based materials. Representative outcomes from these studies are collated in **Figs. 10** and **11**, which highlight the performance of PU and other polymer-derived systems and microcapsules engineered towards dermato-cosmetic applications.

Rheological tests have indicated that enhanced rigidness of the PU foams facilitate their biodegradation. Mudri et al. **[84]** performed shear rheological analyses on Jatropha-oil–derived PU resins, systematically examining the effects of polyol–isocyanate composition alongside standard physicochemical characterizations. Their results revealed distinct flow regimes under ambient conditions, transitioning from Newtonian to shear-thinning (pseudoplastic) behavior depending on the isocyanate content, while both shear-thinning and shear-thickening responses were observed at elevated temperatures. Notably, the authors successfully collapsed rheological data obtained across multiple formulations and temperatures onto a single master curve, thereby enabling model validation and predictive simulation of encapsulation processes (**Fig. 10C**). Complementing these findings, Szpiłyk et al. **[85]** investigated the temperature-dependent shear viscosity (20–80 °C) of biodegradable PU foams synthesized from polyols prepared using glycidol, ethylene carbonate, and cellulose in aqueous media (**Fig. 10D**). Their results highlighted the sensitivity of viscosity to temperature and molecular architecture, reinforcing the importance of polymer composition in governing macroscopic flow properties. Similarly, Parcheta and Datta **[28]** developed fully bio-based linear polyols through a two-step polycondensation reaction and explored their rheology–structure correlation via steady shear rheometry conducted at 60, 70, and 80 °C under controlled shear rates (0–100 $s^{-1}$). The bio-based PU samples exhibited non-Newtonian characteristics consistent with standard rheological models, displaying moderate viscosities (~30 Pa.s) and pronounced pseudoplasticity with minimal yield stress within the examined temperature range. Extending these investigations to functional composites, Wei et al. **[86]** examined the oscillatory rheological response of asphalt systems modified with synthetic PU-based solid–solid phase change materials (PCMs). Using a parallel-plate geometry (25 mm diameter, 1 mm gap) under a strain amplitude of 12% and an angular frequency of 10 $rad \cdot s^{-1}$, they evaluated the complex shear modulus at multiple temperatures (**Figs. 11A** and **11B**). The incorporation of PU–PCM progressively enhanced the deformation resistance and high-temperature stability of the asphalt matrices. Furthermore, a dedicated bending-beam rheological protocol was employed to assess the specimens at sub-ambient temperatures, demonstrating improved low-temperature flexibility and durability.

While these studies involve PU based biomaterials in general, standard rheological studies explicitly involving PU based microcapsules are extraordinarily scarce. Although not strictly PU microcapsules, Naveen et al. [**29**] demonstrated an elaborate strategy to characterize double-walled polyurea based microcapsules (diameter: 300 μm; shell thickness: 1.4 μm) with self-healing features via a rheo-optical setup, using oscillatory shear rheology coupled with a microscope and examining continuous deformations. Specifically, both angular frequency and strain amplitude sweeps were conducted over four orders of magnitudes of strain amplitudes (0.01–100%) and oscillation frequencies (0.1–1000 rad.$s^{-1}$) under varying microcapsule concentrations (5–25%). The authors conformed via rheological studies that incorporation of excipients (epoxy) improved the mechanical properties of double-shelled microcapsules compared to their single-shelled counterparts. On the other hand, recent rheological investigations have also provided insights onto the mechanical behaviour of non-PU based microcapsules, with intended textile and cosmetic applications. Very recently, Tian et al. [**30**] carried out rheological experiments on chitosan-based, self-adhesive microcapsules containing an essential (lavender) oil, prepared via ionic-gelation, and aimed towards functional textile applications. Usual strain amplitude and frequency sweeps were conducted using a parallel plate geometry (diameter: 25 mm; gap: 1 mm) at several orders of magnitudes of the respective parameters and the moduli were recorded under ambient conditions. The microcapsule suspensions showed strain thinning behaviours as a function of varying ingredient mixture. Uniquely, the viscous moduli were higher than their elastic counterparts in the frequency sweeps, irrespective of the combination of the ingredients (**Fig. 11C**). This indicates the influence of the combination of the capsule ingredients on the viscoelastic response. Furthermore, this highlights that association of the capsule components with another polymer, such as guar gum, may alter the molecular network in response to varying degrees of imposed stress deformations. Ali et al. [**87**] reported the fabrication and rheological characterization of silica-graphene oxide crossbreed microcapsules from Pickering emulsions without using surfactants for the regulated release of vanilla oil, a well-known personal care cosmetic ingredient. Oscillatory rheological tests, including oscillation frequency and strain amplitude sweeps were carried out via a cone and plate type geometry (diameter: 40 mm; cone angle 0.3°) over multiple orders of magnitudes, and the viscoelastic moduli were recorded, with predominantly a strain thinning behaviour (**Fig. 11D**). The authors noted that with a gradual increase in the proportion of the silica precursor up to ~20%, the elasticity increases, after which the capsules become rigid, along with a significant increase in the yield stress.

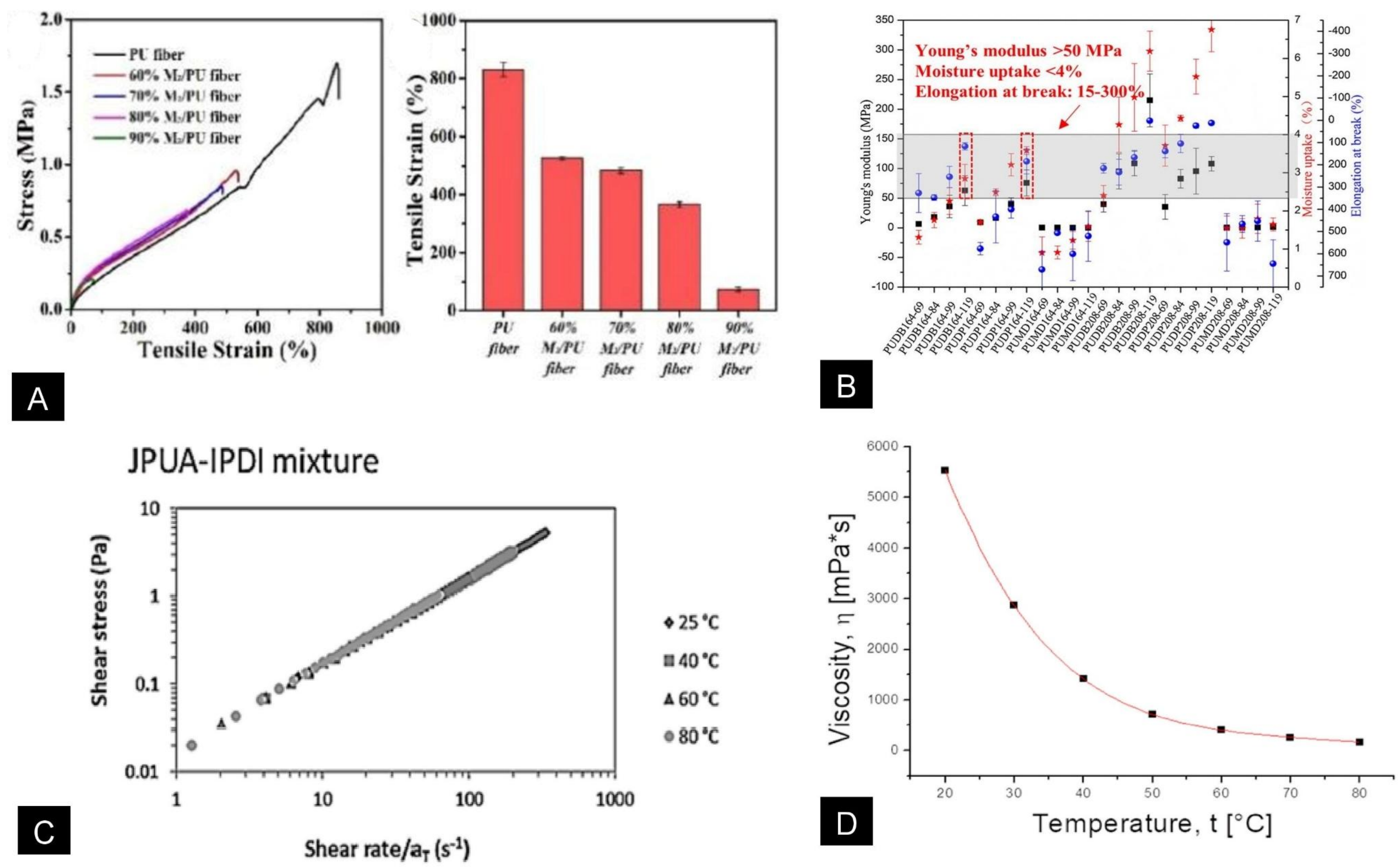

**FIGURE 10.** Mechanical testing, including rheological characterizations of PU/ polymer-based materials / microcapsules towards dermato-cosmetic applications. Part I. **(A):** Tensile stress vs. strain (left) plots and evolution of strain as a function of PU fibre textile / polypyrrole deposited microcapsules (right) at increasing microcapsule concentration (0–90 wt.%), with photothermal energy accommodation and transformation efficacy. Reprinted with permission from Hu et al. [**43**]. Copyright (2021) Elsevier B.V. **(B):** Mechanical characteristics (Young's modulus and elongation at break), and moistness acceptance for bio-based, waterborne PU based films towards haircare applications. Reprinted with permission from Zhang et al. [**56**]. Copyright (2020) Elsevier B.V. **(C):** Master plots of data corresponding to shear rheological characterization of PU resins fabricated from bio-based polyols at different temperatures (25–80 °C), presenting benchmark data for numerical validation. Reprinted from Mudri et al. [**84**]. Open Access. Copyright (2021) The Authors. **(D):** Temperature dependence of shear viscosity for polyols (synthesized from cellulose) towards the fabrication of PU based foams. Reprinted from Szpiłyk et al. [**85**]. Open Access. Copyright (2021) The Authors.

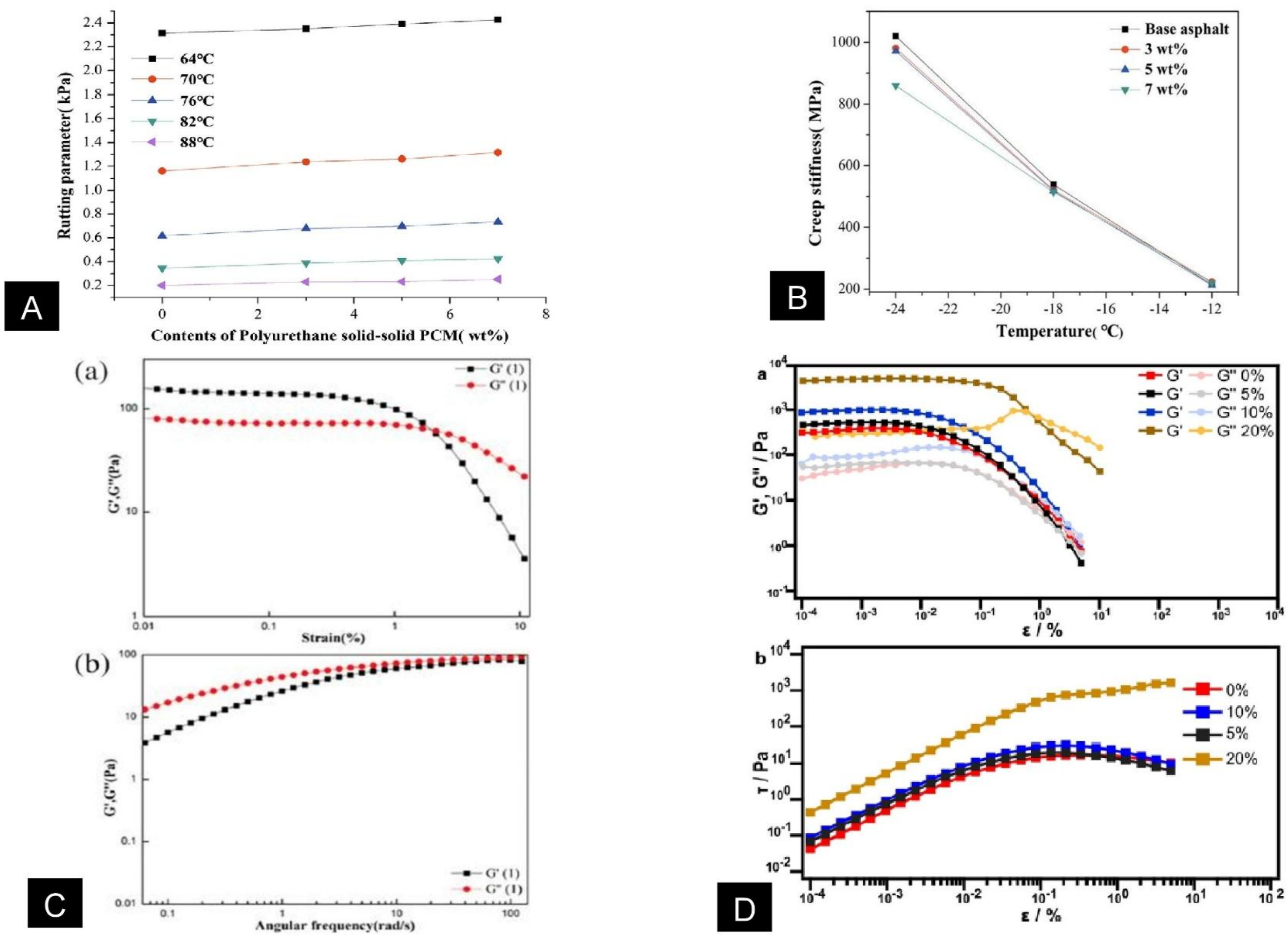

**FIGURE 11.** Mechanical testing, including rheological characterizations of PU/ polymer-based materials / microcapsules towards dermato-cosmetic applications. Part II. **(A):** Rutting parameter (ratio of complex modulus, G* and Sinδ; δ: phase angle) as a function of PU solid-solid PCM towards modified asphalt at different temperatures (60–88 °C). **(B):** Temperature dependence of the creep rigidity as a function of the unmodified and modified asphalt, as in (A), at different proportions of the PU solid-solid PCM. **(A) and (B):** Reprinted with permission from Wei et al. [**86**]. Copyright (2019) International Solar Energy Society. **(C):** Oscillatory rheological testing: strain amplitude sweep showing strain thinning and yielding (a), and angular frequency sweep with predominant viscous behavior at low deformations (b), of guar-gum based self-adhesive microcapsules, with intended applications towards functional textiles. Reprinted with permission from Tian et al. [**30**]. Copyright (2022) Wiley Periodicals LLC. **(D):** Oscillatory rheological testing: strain amplitude sweep (at 1 Hz frequency), showing strain thinning and a weak viscous shear thickening at the highest precursor concentration (20%) (a), and evolution of shear stress as a function of imposed strain amplitude (b), of crossbreed, silica-graphene oxide microcapsules towards personal care cosmetic applications, i.e., for prolonged fragrance discharge. Reprinted with permission from Ali et al. [**87**]. Copyright (2019) Elsevier Inc.

## 4.2. Necessity for exhaustive investigations

### *4.2.1. Addressing the lacunae*

Mechanical testing of microcapsules has been attempted for several years, via tensile testing and mostly focusing on interpreting important parameters, which quantify the degree of elasticity, such as the Young's modulus. For a comprehensive overview, see the very recent review by Xiao et al. [**88**]. However, in recent years, from the discussions highlighting the contemporary state-of-the-art in the previous section, it is apparent that PU based biomaterials, synthesized from bio-based, biocompatible, biodegradable and/or purely synthetic prepolymers, chain extenders, polyols and emulsifiers, were also attempted via both steady state and oscillatory shear deformations. Based on these past studies on PU based hydrogels / scaffolds, films, aerogels, foams, amongst others, preliminary insights onto their mechanical properties were obtained, including knowledge about important parameters like viscosity, viscoelasticity, critical shear rates, strain amplitudes and oscillation frequencies, as well as melting and glass transition temperatures. Despite that, adequate lacunae exist, in terms of rheological characterizations of PU based materials. In this context, it becomes essential to critically assess the current research landscape from the perspective of material specimens. Specifically, while robust and standardized rheological characterization protocols have been established to evaluate the mechanical behavior of polyurethane (PU)-based materials for dermato-cosmetic applications, studies focusing on PU-based microcapsules remain comparatively limited. Moreover, the existing investigations predominantly address their relevance in coatings and adhesive formulations for large-scale industrial applications, such as automotive manufacturing, rather than exploring their potential within dermato-cosmetic domains. This could also be perceived from **Table 1**. This represents a significant knowledge-gap.

### *4.2.2. Characterization strategy*

It is important to evaluate the necessity of the characterization procedure itself, towards the applicability of PU-based microcapsules. To date, majority of the studies has focused on bulk rheological studies under steady state and oscillatory deformations, as a function of multiple orders of magnitudes of operating temperatures, microcapsule concentration in suspension, imposed steady shear rates, angular oscillation frequency and strain amplitudes. These studies have targeted the linear viscoelasticity, temperature and frequency dependence of the storage and loss moduli under small-scale deformations, flow curves as a function of concentration, and shear and complex viscosities. Despite this, severe lacunae persist in the understanding of the behavior of PU microcapsule suspensions under large scale mechanical deformations. For instance, no study has focused on the behavior of PU based microcapsules at shear rates exceeding 1000 $s^{-1}$. Additionally, although a few studies have reported the mechanical response under large-scale oscillatory shear (LAOS) stresses, insights are rather superficial, without additional interpretation via suitable qualitative and quantitative measures of nonlinear viscoelasticity. Since most of the processing in the dermato-cosmetic domains necessitate large-scale material deformations [**6**], relevant characterizations in the nonlinear regime are critical. In this regard, nonexistence of suitably optimized devices could be a vital reason.

#### *4.2.3. Application domain*

When assessing the suitability of PU-based microcapsules, the application domain itself becomes decisive. Critical areas of cosmetic and dermatological research are incomplete without considering the permeation of PCMs or active ingredients, such as vitamins, emollients, and essential oils. Additionally, the fields are also associated with perspectives. For instance, while ingredient concentration and sustained release warrant attention from a host interaction perspective, large-scale processing operations, such as extrusion and molding, becomes crucial from a manufacturing perspective. Furthermore, the end-product, based on representative application, is also important. For instance, applications of PU based microcapsules towards haircare and lipcare products would require rigorous investigations and interpretation of the large-scale stress response at very high deformations. To the best of our knowledge, systematic rheological investigations under large-scale mechanical deformations of PU based microcapsules, bio-based or synthetic, and a subsequent interpretation of the important parameters, are virtually nonexistent in literature. This concerns both the critical stress for the break-up of microcapsules, as well as nonlinear viscoelastic or flow-based behaviors far from equilibrium, while targeting skin therapeutics, wearable and personal care cosmetic applications. Microrheological studies, including accompanying optical and microfluidic techniques, are out of scope for the current review.

## 4.3. Target domains

Due to high necessity, we deliberate several distinct target areas for in-depth rheological characterizations. The discussions below are exclusively aimed at PU based microcapsules, with targeted dermato-cosmetic applications.

### *4.3.1. Microcapsule ingredients and configurations*

A comprehensive investigation is required to elucidate how the rheological responses of PU microcapsules, derived from both bio-based and synthetic prepolymers, chain extenders, and isocyanates, vary under low- to medium steady shear and small-amplitude oscillatory shear (SAOS) deformations. This includes varying degrees of influence exerted by cross-linking agents, as well as synthetic emulsifiers and their bio-based counterparts, such as rhamnolipids and Eco-Tween 80, including their concentrations and type. Additionally, whether and how far the structural conformations affect the mechanical response. For instance, is there any difference in the response between core-shell type PU microcapsules and PU based microspheres? Also, in the context of core-shell capsules, whether mononuclear with double-walled capsules show a varied response to mechanical stress compared to a polynuclear structure with a single wall.

**TABLE 1.** Dermato-cosmetic applications of PU based materials: nature of PU (based on the prepolymer / chain extender / polyol), proposed or demonstrated applications, type of PU based material, representative characteristics (shape / size / size range in diameter), mechanical testing methods, rheological characterizations (type of deformation and relevant specifics), and unique attributes associated with PU incorporation. Abbreviations: PU – polyurethane, MC – microcapsule, SS – strain amplitude sweep, FS – frequency sweep, PCM – phase change material, N.A. – not applicable.

| **Nature of PU** | **Applications** | **Type of PU based material** | **MC morphology and size distribution (diameter)** | **Mechanical testing** | **Rheological characterizations (relevant specifics)** | **Remarks** | **Reference** |
|---|---|---|---|---|---|---|---|
| Synthetic | Tissue engineering | Self-healing injectable hydrogels | N.A. | Via oscillatory shear rheology | SS (0.01–5000% strain); FS (20–100 rad.$s^{-1}$); complex viscosity profiles (5–60 °C) | 20-fold increase in gel compression strength | Balavigneswaran and Muthuvijayan [78] |
| Synthetic | Chronic diabetic wound healing | Nanofibrous scaffolds | N.A. | Via measuring tensile strengths and elongation at break | N.A. | First time synthesis of PU / polyvinyl-alcohol based nanofibrous membranes | Mohamady Hussein et al. [79] |
| Partially bio-based (aliphatic isocyanates) | Drug delivery | Shape memory microcapsules | Spherical, ~2-10μm | Via tensile stretching and customized quantitative techniques | N.A. | A general, scalable method to produce shape memory PU based MC | Zhang et al. [19] |
| Partially bio-based (Gum Arabic) | Antibacterial / antifouling agents | Double-layer, core-shell type microcapsules | Spherical, 102.2 ± 17.3μm | N.A. | N.A. | Antimicrobial actions against marine and marine biofilm forming bacteria | Chong et al. [20] |
| Synthetic | Implant design / self-healing coatings | Core-shell type microcapsules | Spherical, 50-720μm | N.A. | N.A. | Anti-corrosion performance positively correlated with increasing MC loading | Parsaee et al. [23] |
| Synthetic | Cosmetotextiles | Core-shell type microcapsules | Spherical, ~3-50μm | N.A. | N.A. | MCs stable up to 4 washes on fabric. High retention on silk. | Mertgenç et al. [17] |
| Synthetic | Cosmetics | Double-walled, core-shell type microcapsules | Spherical, ~3-5μm, smooth and compact surface | N.A. | N.A. | Double-walled MCs showed higher thermal stability and firmness than single-walled MCs | Lu et al. [26] |
| Partially bio-based (isosorbide/biomass derived) | Cosmetotextiles | Core-shell type microcapsules | Spherical, ~2-100μm, rough surface | N.A. | N.A. | MCs detected on cotton fabric even after 40 washing cycles | Ben Abdelkader et al. [50] |
| Bio-based | Cosmetics, hair styling agents | Films | N.A. | Via measuring tensile strengths and elongation at break | N.A. | Low moisture uptake rate, decent mechanical features, no indication of skin irritation | Zhang et al. [56] |
| Synthetic | Textiles | Electrospun nanofibrous membranes | N.A. | N.A. | N.A. | Enhanced thermal insulation via PU membrane embedding to silica aerogel | Venkataraman et al. [42] |
| Synthetic | Textiles | Core-shell type microcapsules | Spherical, size range unspecified | N.A. | N.A. | Enhanced thermal performance via PU-modified MC shells | Skurkyte-Papieviene et al. [12] |
| Partially bio-based ($\beta$-cyclodextrin based) | Textiles | Core-shell type microcapsules | Spherical, ~1-40μm | N.A. | N.A. | Loaded MCs preserved competence even after 35 washing cycles | Azizi et al. [21] |

| Bio-based linear polyester polyols (towards PU synthesis) | Any proposed PU application | Unspecified | N.A. | Via steady shear rheology | Viscosity and flow curves (0–100 $s^{-1}$); Rotary, cone/plate geometry. Dynamic viscosity under controlled shear rates | Non-Newtonian behavior. Pseudoplasticity with lowest yield stress at 60, 70 and 80 °C. | Parcheta and Datta [28] |
|---|---|---|---|---|---|---|---|
| Bio-based (via cellulose-based polyols) | Any proposed PU application | Foam | N.A. | Via steady shear rheology | Temperature sweep (20–80 °C) of shear viscosity. | High thermal resistance (up to 150 °C). Polyol 100% and PU foam 70-80% biodegradable after 28 days. | Szpiłyk et al. [85] |
| Synthetic (towards modified asphalt synthesis) | Self-healing, anti-aging agent | Solid-solid PCM | N.A. | Via dynamic shear rheology and bending-beam rheology | Complex modulus ($G^*$) and phase angle ($\delta$) as a function of PU solid-solid PCM (0–8 wt.%). Temperature dependence (-24, -18 and -12 °C) of the creep stiffness. | Enhanced high-temperature performance of asphalt via incorporation of PU solid-solid PCM | Wei et al. [86] |
| Partially bio-based (vegetable oil based) polyols (towards PU acrylate synthesis) | Self-healing coatings, medical storage and processing | Resin | N.A. | Via steady shear rheology | Viscosity and flow curves (0–100 $s^{-1}$) at different temperatures (25, 40 60, and 80 °C) | Shear-rate based master plots of shear stress data at different temperatures | Mudri et al. [84] |

### *4.3.2. Operating conditions and characterization setups*

It remains to be established whether and to what extent, principal operating parameters such as temperature, as well as the type of geometry of the rheometer significantly affect the mechanical response of the PU microcapsule suspensions. This can be investigated, for instance, by correlating the thermal stability data from a different procedure, such as TGA or DSC, with that of bulk temperature dependence of either the shear viscosity or the linear viscoelastic moduli. It is apparent that microcapsule suspensions can be well-characterized with a 'cup-and-bob' or concentric cylinder geometries facilitating a Couette flow [**27**], imposing steady shear deformations. Data from other geometries, i.e., parallel-plate and cone-and-plate, could offer valuable additional insights to this, especially towards their oscillatory response. Furthermore, whether problems such as wall-slip could be avoided just by using a rough surface rather than a smooth one, as is the case with entangled polymer solutions, gels and melts.

### *4.3.3. Concentration dependent dynamics at high steady shear deformations*

The concentration dependent dynamics of PU microcapsule suspensions towards intended dermato-cosmetic applications would be a valuable addition to the literature. The shear viscosity of the suspensions can now be measured at extended orders of magnitudes of shear rates, resulting in complete viscosity flow curves (see section 5.3). From very low to critical shear stress, these will highlight distinct aspects of relevant dermato-cosmetic investigations corresponding to different stages, such as storage (steady-state), extrusion (medium to high shear deformations), application on skin (high shear deformations), and spreadability and assimilation (yielding). Furthermore, topical applications of several cosmetic products, such as lipsticks, necessitate rheological characterizations at high shear rates, i.e., $10^3$-$10^4$ $s^{-1}$, to interpret their physiological relevance [**41**]. No study to date has reported the behavior of infinite shear viscosity and/or normal stress differences of microcapsule-based suspensions towards dermato-cosmetic applications, at such high shear rates. Apparently, despite high necessity, a systematic study of shear stress tolerance of PU microcapsules as a function of parameters such as operating temperature, characterization geometry, suspending solvent, functional additives, amongst others, is still lacking in the literature.

### *4.3.4. Concentration dependent dynamics under LAOS stresses*

Likewise, oscillatory shear response of PU microcapsule suspensions from low to high concentrations under LAOS stresses will reveal real material properties in the nonlinear deformation range. For instance, a host of relevant viscoelastic parameters, such as tangent and secant moduli at the maximum and minimum strain amplitudes or alternatively, at minimum and maximum shear rates, could be interpreted in addition to the usual linear viscoelastic moduli, in response to wound healing, oralcare and haircare applications. For instance, it is not yet established how extrusion design for three-dimensional (3D) and four-dimensional (4D) bioprinting of novel tissue engineering scaffolds as well as medical implant devices could serve as an ideal application area for the PU based microcapsules. 3D printing constructs three-dimensional, stationary assemblies or configurations through a layer-by-layer deposition approach, guided by a digital or numerical model **[89]**. In contrast, 4D printing employs a similar additive process but integrates 'smart' materials capable of altering their forms or a self-assembling approach, mostly triggered via peripheral stimuli such as humidity, light or temperature **[90]**. Fundamentally, 4D printing incorporates the temporal dimension into the synthesis process, enabling the manufacturing of dynamic, adaptable objects, rather than immobile, fixed ones obtained via 3D printing. However, this could be validated to a large extent based on the nonlinear viscoelastic parameters. This also facilitates a dynamic examination of the complex association between microcapsule concentration, yielding, thixotropy and microstructure, in great detail.

### *4.3.5. Qualitative and quantitative interpretations of LAOS data*

LAOS interpretation of nonlinear viscoelastic data corresponding to a wide range of specimens, i.e., polymer and colloid based biomaterials such as scaffolds, gels, suspensions, amongst others, has received widespread attention [**40, 91**; and references therein]. Despite that, standard and unique modes of interpretation of PU microcapsule LAOS data could be very enticing for prospective dermato-cosmetic applications, under static or steady state conditions, as well as a time-dependent dynamic response. For instance, in drug delivery studies, syringe handling and subcutaneous delivery of injectable self-healing PU based biomaterials including vesicular compartments, application of cosmetic agents with encapsulated active ingredients or PCMs on the host skin, as well as regulated manipulation of capsules or microspheres in miniature lab-on-a-chip nanofluidic or microfluidic devices, would necessitate interpretation of rheological or viscoelastic response under nonlinear stress deformations. In this regard, exhaustive investigations of standard nonlinear rheological data interpretation tools, including Lissajous-Bowditch curves, third-order Chebyshev coefficients, dimensionless strain-thinning and shear-thickening coefficients, amongst others, will shed light on the intracycle and intracycle viscoelastic response. Contextually, effective case studies were documented recently towards innovative dermato-cosmetic applications [**38, 40, 41**].

#### *4.3.6. Viscoelastic properties via creep-ringing studies*

A few recent creep studies have focused on shape memory or self-assembly targeted PU microcapsules towards either improving the recovery ratios of the capsules or delaying the gel gelation time [**73, 92**]. These studies show the efficacy of creep studies in obtaining tangible viscoelastic data for dense PU microcapsule suspensions. One would argue that there are only limited physiological applications of such suspensions in areas such as drug delivery. Nevertheless, in the areas of scaffold design and therapeutic coatings, and in cosmetotextiles, stress dependent yielding would be an interesting domain of study. Consequently, viscoelastic properties of such high concentration PU microcapsule suspensions can be determined via creep-ringing under varying degrees of imposed stresses, including the development of shear rates during ringing. The creep ringing technique is particularly helpful, since it is possible to interpret the viscoelastic characteristics as well as the long-duration creep compliance data simultaneously. Moreover, benchmark experimental data can also be validated via standard, multiple-element models, such as the Jeffreys model.

### *4.3.7. Associating rheological data with sensory and permeation characteristics*

Data handling after rheological measurements is equally crucial. Clearly, the structure function relationships critically govern the development of innovative cosmetic products. On the one hand, this can provide insightful behavior of the underlying composites, wax bases, drug formulations, organogels, implant and scaffold matrices, being subject to varying degrees of deformations. On the other hand, useful measures that manifest the sensory profiles of such products, including quick break, cushion, slip, amongst others, can be associated with the rheological data, and could be further developed via examination by trained personnel. For instance, in composite skin care formulations, the zero-shear rate viscosity from steady flow curve, and the stress maxima from strain amplitude sweep, have been shown to be effective indicators of cushion and easy break sensory characteristics [**41**; and references therein). In addition, *in vitro* permeation or release data via diffusion tests of active dermato-cosmetic ingredients, such as vitamin D3 and caffeine, have been shown to actively validate with rheological measurements [**2, 38**]. Subsequently, this could also be applied to the PCMs in the PU microcapsules.

### *4.3.8. Optimization towards domain-specific investigations*

A vast majority of the prospective rheological assessments has to be comprehended and optimized with respect to the target applications. For instance, the most appropriate rheological characterization towards high concentration PU microcapsule containing protein-drug formulation would be to carry out steady shear deformations at high shear rates, mainly to address the well-known ‘syringe problem’, i.e., how to overcome the interfacial and wall effects of an injection syringe during subcutaneous penetration while minimizing protein aggregation. A side investigation would be to examine the direct effects of shear on the aggregation clusters. From another viewpoint, most personal care cosmetic merchandises, such as lipcare and haircare products, necessitate data on their viscoelastic properties under small to large scale mechanical deformations, thereby requiring oscillatory rheological characterizations. In domains such as cosmetotextiles, tissue engineering, therapeutic coatings and adhesives, PU microcapsule comprising scaffolds, membranes and thin films would necessitate stress relaxation, yield stress and creep ringing experiments.

## 5. CHALLENGES, VIEWPOINT AND OPPORTUNITIES

### 5.1. Focus on bio-based PU microcapsules

A careful introspection of the current state-of-the-art reveal two broad progress-oriented routes concerning both the synthesis of PU based microcapsules, as well as their rheological characterizations and intended applications in the dermato-cosmetic sectors. On the one hand, PU based biomaterials have found recent applications in the aforementioned sectors. On the other hand, although microcapsules based on other polymers have found applications in these domains, PU based microcapsule applications are extraordinarily scarce. One primary reason could be the changing perception of customers, associated with propensity towards more natural component based 'green' cosmetics and dermatological products, which necessitates an increasing utilization of bio-based PU microcapsules. This is currently not the case, since majority of the PU based applications in cosmetics deals with synthetic PU [**8**]. The challenge is to synthesize either microspheres or core-shell type microcapsules with either single, double or even multi-shells with bio-based PU as one of the components, which would suit and can be optimized depending on the domain. Here intercomponent compatibility is the key factor. For this, opportunities lie in employing exclusively natural pre-polymers, chain extenders and isocyanates towards PU synthesis. For instance, the work done in recent years by Loos and coworkers [**93, 94**] focusing on fabricating essentially 'green' PU from renewable isocyanates and bio-based white dextrins, holds promise. Contextually, Lu et al. [**26**] reported the PU and polyurea based microcapsules with enhanced thermal stability, which could have potential cosmetic applications. However, it is not straightforward to assert that these natural components would suffice and generate the desired PU microcapsules, which could also demonstrate desired mechanical response via a wide range of rheological tests, as discussed in the previous section. Nonetheless, this could be an interesting case study.

### 5.2. Customized, domain-tailored applications

Shortcomings and hindrances pertaining to the applications of PU based microcapsules have also been highlighted customized to tailored domains. For instance, a range of imminent impediments has been identified by Sobczak and Kędra [**22**] in PU based drug delivery schemes, specifically to counter carcinogenesis, including undesired host immunological response, predominant usage of expensive and synthetic starting materials towards their fabrication, exhaustive multi-stage synthesis procedures, impulsive and abrupt release of PCMs, and adverse interactions with certain biomolecules catering to their large surface areas. On the bright side, efficacy of double-shelled PU based microcapsules as a potential substitute towards manufacturing of a sustainable, antifouling agents has been highlighted, subject to careful manipulation of the choice of PCM (such as clove oil), as well as the proportion of the PU components [**20**]. This presents a clear opportunity to test the same towards potential dermato-cosmetic applications.

### 5.3. Mechanical features via original rheological measurement techniques

Additionally, from the purview of rheological characterizations, in addition to the target domains discussed earlier, a significant challenge lies in the context of the choice of rheometer itself. For instance, commercially available rotational rheometers, either strain or stress controlled, can typically access shear rates in the low to medium shear rate range ($\leq 10^3$ $s^{-1}$). However, effective characterizations simulating physiological applications of certain composite cosmetic formulations, including lipsticks and hairgels, typically require measurements at shear rates at around $10^4$ $s^{-1}$ or higher [**41, 95**]. Consequently, this presents an opportunity to optimize existing devices catering to the applications. Contextually, Wierschem and coworkers [**96-99**] have demonstrated the efficacy of a customized narrow-gap rotational rheometer, which can access shear rates of ~$10^5$ $s^{-1}$. Undoubtedly, this offers prospects of characterizing either thin-film textiles embedded with PU microcapsules, and typical PU microcapsule loaded cosmetic formulations at ultra-high shear rates, simulating manifestations embodying yielding (stress tolerance), spreading (on the skin), amongst others.

### 5.4. Tangible scale up prospects

An often overlooked, but otherwise critically important aspect is the scale up opportunities associated with the development of any commercial product [**100**] including cosmetic formulations. Like several other fundamental areas of microcapsule applications, cosmeto-dermatological research has a devoted projection towards industrial processability. Hence, it is crucial not to overestimate the outcomes derived solely from lab-scale investigations, while facilitating and pivoting the initial successful results towards a tangible, pilot scale operation. This is also applicable for the PU based microcapsules. To this end, the study by Zhang et al. [**19**] holds promise, who have highlighted the scale-up potential of their general PU shape memory microcapsules, manufactured via aliphatic triisocyanates, mainly towards drug delivery.

### 5.5. Customer-focused applications and legislations

Optimizing applications pertaining to the target customer base, is a perpetual endeavor, especially in the cosmetics sector. Furthermore, customer preference is a broad term, transcending critical avenues such as anthropology, ethos, micro- and macroclimate, geographical location, and above all, sensitivity and mood. For instance, individuals living in perpetually colder regions would most-likely prefer a more wearable PU based product with a sustained product release, and hence, products with much higher PU based microcapsule concentration compared to ones thriving in a tropical climate. From a different perspective, PU microcapsules targeted towards the potential delivery of an active pharmaceutical ingredient like vitamin D3 would have to be incorporated at different proportions in the intended products, depending upon customer ethnicity, citing the differential absorbance of vitamin D3 depending upon the skin melanin content. Finally, legislations and regulations attributed to specific geographical locations, countries or alliances, should be strictly complied with, while trying to formulate novel PU microcapsule incorporated products. This may either restrict or encourage the usage of certain ingredients or testing of certain end products, for instance banned animal testing of cosmetic raw materials or utilization of borderline products in the EU region [**101**], which need to be enforced.

## 6. CONCLUSIONS

In the past 7 to 8 years, we note an upsurge of studies targeting polyurethane-based microencapsulation applications towards dermato-cosmetic interventions. Apart from the intrinsic benefits of polyurethane itself and of microencapsulation technology, this trend has witnessed a broader macroeconomic progress, a transformation of global customer outlook, and a penchant for more bio-based, bio-compatible and biodegradable choice of raw materials and ingredients. These drifts also lay support on the benefits of the microencapsulation technology, from the perspectives of capsule configurations, efficient case-driven selection of the phase change materials, and choice of the next generation of emulsifiers that minimize hazards but enhance optimization. These, in turn, facilitate novel designs of polyurethane based microencapsulated vehicles towards dermato-cosmetic applications. As much for the advantages, we also note severe limitations of the technology, especially in mechanical characterization. Specifically, understanding and interpretation of mechanical response only based on traditional protocols rigorously limits our understanding. In this regard, we highlight the necessity of advanced, bulk rheological characterizations of polyurethane-based microcapsule suspensions in dermato-cosmetic applications, based on a wide range of viewpoints, highlighting applicability, procedures, shortcomings and challenges. In this regard, we foresee a significant advancement in the applicability of polyurethane-based microcapsules in the coming years, but not without strict appraisals enforced, as well as appropriate and rational steps undertaken to tackle the deliberated challenges.

## CRediT authorship contribution statement

**Sharadwata Pan:** Conceptualization, Method, Validation, Formal analysis, Investigation, Resources, Writing – Original draft. Writing **–** review & editing. **Andreas Wierschem:** Writing – Review & editing. **Natalie Germann:** Funding acquisition, Writing **–** review & editing. **Thomas Becker:** Supervision, Project administration.

## ACKNOWLEDGEMENTS

SP was supported by the *Arbeitsgemeinschaft industrieller Forschungsvereinigungen* „Otto von Guericke“ e.V. (AiF) Grant (Funding code: KK 5022303SK0; AiF Projekt GmbH), a strategic cooperation between the Technical University of Munich and ODVITAL Cosmetics GmbH, as a part of the Central Innovation Program (*ZIM-Kooperationsprojekt*) Initiative of the Federal Ministry for Economic Affairs and Energy, Federal Government of Germany. We acknowledge the feedback and suggestions by the anonymous reviewers, which improved the quality of the manuscript.

## DECLARATION OF COMPETING INTEREST

All authors declare no competing financial interests and personal relationships with other people or organizations that could have inappropriately influenced (biased) our work.

## DATA AVAILABILITY

No data was used for the research described in the article.

## REFERENCES

1. Suphasomboon, T., Vassanadumrongdee, S. Toward sustainable consumption of green cosmetics and personal care products: The role of perceived value and ethical concern (2022) Sustainable Production and Consumption, 33, pp. 230-243. DOI: 10.1016/j.spc.2022.07.004

2. Pan, S., Sivanathan, S., Kiepe, G., Kiepe, T., Germann, N. Candidate Formulations for a Sustainable Lipstick Supplemented with Vitamin D3: Effects of Wax Type and Concentration on Material Properties (2021a) Industrial and Engineering Chemistry Research, 60 (5), pp. 2027-2040. DOI: 10.1021/acs.iecr.0c05264

3. Meftahi, A., Samyn, P., Geravand, S.A., Khajavi, R., Alibkhshi, S., Bechelany, M., Barhoum, A. Nanocelluloses as skin biocompatible materials for skincare, cosmetics, and healthcare: Formulations, regulations, and emerging applications (2022) Carbohydrate Polymers, 278, art. no. 118956,. DOI: 10.1016/j.carbpol.2021.118956

4. Pan, S., Germann, N. Thermal and mechanical properties of industrial benchmark lipstick prototypes (2019) Thermochimica Acta, 679, art. no. 178332,. DOI: 10.1016/j.tca.2019.178332

5. Dini, I., Laneri, S. The new challenge of green cosmetics: Natural food ingredients for cosmetic formulations (2021) Molecules, 26 (13), art. no. 3921,. DOI: 10.3390/molecules26133921

6. Pan, S., Goudoulas, T.B., Jeevanandam, J., Tan, K.X., Chowdhury, S., Danquah, M.K. Therapeutic Applications of Metal and Metal-Oxide Nanoparticles: Dermato-Cosmetic Perspectives (2021b) Frontiers in Bioengineering and Biotechnology, 9, art. no. 724499,. DOI: 10.3389/fbioe.2021.724499

7. Gebashe, F.C., Naidoo, D., Amoo, S.O., Masondo, N.A. Cosmeceuticals: A Newly Expanding Industry in South Africa (2022) Cosmetics, 9 (4), art. no. 77,. DOI: 10.3390/cosmetics9040077

8. Alves, T.F.R., Morsink, M., Batain, F., Chaud, M.V., Almeida, T., Fernandes, D.A., da Silva, C.F., Souto, E.B., Severino, P. Applications of natural, semi-synthetic, and synthetic polymers in cosmetic formulations (2020) Cosmetics, 7 (4), art. no. 0075,. DOI: 10.3390/COSMETICS7040075

9. Perinelli, D.R., Palmieri, G.F., Cespi, M., Bonacucina, G. Encapsulation of Flavours and Fragrances into Polymeric Capsules and Cyclodextrins Inclusion Complexes: An Update (2020) Molecules, 25 (24), art. no. 5878,. DOI: 10.3390/MOLECULES25245878

10. Ma, K., Zhang, X., Ji, J., Han, L., Ding, X., Xie, W. Application and research progress of phase change materials in biomedical field (2021) Biomaterials Science, 9 (17), pp. 5762-5780. DOI: 10.1039/d1bm00719j

11. Xiao, Z., Liu, H., Zhao, Q., Niu, Y., Chen, Z., Zhao, D. Application of microencapsulation technology in silk fibers (2022) Journal of Applied Polymer Science, 139 (25), art. no. e52351,. DOI: 10.1002/app.52351

12. Skurkyte-Papieviene, V., Abraitiene, A., Sankauskaite, A., Rubeziene, V., Baltusnikaite-Guzaitiene, J. Enhancement of the thermal performance of the paraffin-based microcapsules intended for textile applications (2021) Polymers, 13 (7), art. no. 1120,. DOI: 10.3390/polym13071120

13. Petrulis, D., Petrulyte, S. Potential use of microcapsules in manufacture of fibrous products: A review (2019) Journal of Applied Polymer Science, 136 (7), art. no. 47066,. DOI: 10.1002/app.47066

14. Boh Podgornik, B., Šandric, S., Kert, M. Microencapsulation for functional textile coatings with emphasis on biodegradability—a systematic review (2021) Coatings, 11 (11), art. no. 1371,. DOI: 10.3390/coatings11111371

15. De Souza F.M., Kahol P.K., Gupta R.K. Introduction to Polyurethane Chemistry (2021) ACS Symposium Series, 1380, pp. 1 – 24. DOI: 10.1021/bk-2021-1380.ch001

16. Ben Abdelkader, M., Azizi, N., Baffoun, A., Bordes, C., Chevalier, Y., Majdoub, M. Application of an experimental design methodology for the optimization of cosmetotextile impregnation process by β-cyclodextrin based microcapsules (2022) Journal of the Textile Institute,. DOI: 10.1080/00405000.2022.2031710

17. Mertgenç, C., Enginar, H., Yılmaz, H. Microencapsulation of Fragrance with Polyurethane—Urea and Application on Different Fabrics (2021) Iranian Journal of Science and Technology, Transaction A: Science, 45 (5), pp. 1677-1687. DOI: 10.1007/s40995-021-01135-y

18. Jeevanandam, J., Pan, S., Danquah, M.K. Biomedical and Environmental Applications of Waterborne Polyurethane-Metal Oxide Nanocomposites (2021) Advances in Science, Technology and Innovation, pp. 179-192. DOI: 10.1007/978-3-030-72869-4_12

19. Zhang, F., Zhao, T., Ruiz-Molina, D., Liu, Y., Roscini, C., Leng, J., Smoukov, S.K. Shape Memory Polyurethane Microcapsules with Active Deformation (2020a) ACS Applied Materials and Interfaces, 12 (41), pp. 47059-47064. DOI: 10.1021/acsami.0c14882

20. Chong, Y.-B., Zhang, H., Yue, C.Y., Yang, J. Fabrication and Release Behavior of Microcapsules with Double-Layered Shell Containing Clove Oil for Antibacterial Applications (2018) ACS Applied Materials and Interfaces, 10 (18), pp. 15532-15541. DOI: 10.1021/acsami.8b05467

21. Azizi, N., Ben Abdelkader, M., Chevalier, Y., Majdoub, M. New β-Cyclodextrin-Based Microcapsules for Textiles Uses (2019) Fibers and Polymers, 20 (4), pp. 683-689. DOI: 10.1007/s12221-019-7289-5

22. Sobczak, M., Kędra, K. Biomedical Polyurethanes for Anti-Cancer Drug Delivery Systems: A Brief, Comprehensive Review (2022) International Journal of Molecular Sciences, 23 (15), art. no. 8181,. DOI: 10.3390/ijms23158181

23. Parsaee, S., Mirabedini, S.M., Farnood, R., Alizadegan, F. Development of self-healing coatings based on urea-formaldehyde/polyurethane microcapsules containing epoxy resin (2020) Journal of Applied Polymer Science, 137 (41), art. no. 49663,. DOI: 10.1002/app.49663

24. Zhou, X., Zhang, X., Mengyuan, P., He, X., Zhang, C. Bio-based polyurethane aqueous dispersions (2021) Biopolymers and Composites: Processing and Characterization, pp. 155-188. DOI: 10.1515/9781501521942-005

25. Yin, Q., Guo, M., Li, W., Ren, H., Wang, J., Zhang, X. Fabrication and Performance of Polyurethane / Polyurea Microencapsulated Phase Change Materials with Isophorone Diisocyanate via Interfacial Polymerization (2019) Science of Advanced Materials, 11 (5), pp. 647-654(8). DOI: 10.1166/sam.2019.3407

26. Lu, S., Shen, T., Xing, J., Song, Q., Xin, C. Preparation, characterization, and thermal stability of double-composition shell microencapsulated phase change material by interfacial polymerization (2017) Colloid and Polymer Science, 295 (10), pp. 2061-2067. DOI: 10.1007/s00396-017-4189-3

27. Cao, V.D., Salas-Bringas, C., Schüller, R.B., Szczotok, A.M., Hiorth, M., Carmona, M., Rodriguez, J.F., Kjøniksen, A.-L. Rheological and thermal properties of suspensions of microcapsules containing phase change materials (2018) Colloid and Polymer Science, 296 (5), pp. 981-988. DOI: 10.1007/s00396-018-4316-9

28. Parcheta, P., Datta, J. Structure-rheology relationship of fully bio-based linear polyester polyols for polyurethanes - Synthesis and investigation (2018) Polymer Testing, 67, pp. 110-121. DOI: 10.1016/j.polymertesting.2018.02.022

29. Naveen, V., Raja, S., Deshpande, A.P. A two-step approach for synthesis, characterization and analysis of dicyclopentadiene–urea formaldehyde–siloxane-based double-walled microcapsules used in self-healing composites (2019) International Journal of Plastics Technology, 23 (2), pp. 157-169. DOI: 10.1007/s12588-019-09247-2

30. Tian, Y., Huang, X., Cheng, Y., Niu, Y., Meng, Q., Ma, J., Zhao, Y., Kou, X., Ke, Q. Preparation of self-adhesive microcapsules and their application in functional textiles (2022) Journal of Applied Polymer Science, 139 (29), art. no. e52650,. DOI: 10.1002/app.52650

31. Pan, S., Nguyen, D.A., Sunthar, P., Sridhar, T., Prakash, J.R. Uniaxial extensional viscosity of semidilute DNA solutions (2019b) Korea Australia Rheology Journal, 31 (4), pp. 255-266. DOI: 10.1007/s13367-019-0026-1

32. Tafuro, G., Costantini, A., Baratto, G., Francescato, S., Busata, L., Semenzato, A. Characterization of polysaccharidic associations for cosmetic use: Rheology and texture analysis (2021) Cosmetics, 8 (3), art. no. 62,. DOI: 10.3390/cosmetics8030062

33. Pan, S., Nguyen, D.A., Dünweg, B., Sunthar, P., Sridhar, T., Ravi Prakash, J. Shear thinning in dilute and semidilute solutions of polystyrene and DNA (2018) Journal of Rheology, 62 (4), pp. 845-867. DOI: 10.1122/1.5010203

34. Kokkinos D., Dakhil H., Wierschem A., Briesen H., Braun A. Deformation and rupture of Dunaliella salina at high shear rates without the use of thickeners (2016) Biorheology, 53 (1), pp. 1 – 11. DOI: 10.3233/BIR-15057

35. Dakhil H., Gilbert D.F., Malhotra D., Limmer A., Engelhardt H., Amtmann A., Hansmann J., Hübner H., Buchholz R., Friedrich O., Wierschem A. Measuring average rheological quantities of cell monolayers in the linear viscoelastic regime (2016) Rheologica Acta, 55 (7), pp. 527 – 536. DOI: 10.1007/s00397-016-0936-5

36. Dakhil H., Do H., Hübner H., Wierschem A. Measuring the adhesion limit of fibronectin for fibroblasts with a narrow-gap rotational rheometer (2018) Bioprocess and Biosystems Engineering, 41 (3), pp. 353 – 358. DOI: 10.1007/s00449-017-1868-x

37. Lee S., Bashir K.M.I., Jung D.H., Basu S.K., Seo G., Cho M.-G., Wierschem A. Measuring the linear viscoelastic regime of MCF-7 cells with a monolayer rheometer in the presence of microtubule-active anti-cancer drugs at high concentrations (2022) Interface Focus, 12 (6), art. no. 20220036. DOI: 10.1098/rsfs.2022.0036

38. Malhotra, D., Pan, S., Rüther, L., Schlippe, G., Voss, W., Germann, N. Polysaccharide-based skin scaffolds with enhanced mechanical compatibility with native human skin (2021) Journal of the Mechanical Behavior of Biomedical Materials, 122, art. no. 104607,. DOI: 10.1016/j.jmbbm.2021.104607

39. Malhotra, D., Pan, S., Rüther, L., Goudoulas, T.B., Schlippe, G., Voss, W., Germann, N. Linear viscoelastic and microstructural properties of native male human skin and in vitro 3D reconstructed skin models (2019) Journal of the Mechanical Behavior of Biomedical Materials, 90, pp. 644-654. DOI: 10.1016/j.jmbbm.2018.11.013

40. Pan, S., Malhotra, D., Germann, N. Nonlinear viscoelastic properties of native male human skin and in vitro 3D reconstructed skin models under LAOS stress (2019a) Journal of the Mechanical Behavior of Biomedical Materials, 96, pp. 310-323. DOI: 10.1016/j.jmbbm.2019.04.032

41. Pan, S., Germann, N. Mechanical response of industrial benchmark lipsticks under large-scale deformations (2020) Acta Mechanica, 231 (7), pp. 3031-3042. DOI: 10.1007/s00707-020-02691-x

42. Venkataraman, M., Mishra, R., Militky, J., Xiong, X., Marek, J., Yao, J., Zhu, G. Electrospun nanofibrous membranes embedded with aerogel for advanced thermal and transport properties (2018) Polymers for Advanced Technologies, 29 (10), pp. 2583-2592. DOI: 10.1002/pat.4369

43. Hu, L., Li, X., Ding, L., Chen, L., Zhu, X., Mao, Z., Feng, X., Sui, X., Wang, B. Flexible textiles with polypyrrole deposited phase change microcapsules for efficient photothermal energy conversion and storage (2021) Solar Energy Materials and Solar Cells, 224, art. no. 110985,. DOI: 10.1016/j.solmat.2021.110985

44. De Castro, P.F., Minko, S., Vinokurov, V., Cherednichenko, K., Shchukin, D.G. Long-Term Autonomic Thermoregulating Fabrics Based on Microencapsulated Phase Change Materials (2021) ACS Applied Energy Materials, 4 (11), pp. 12789-12797. DOI: 10.1021/acsaem.1c02170

45. Qureshi S.A., Dorugade V.A., Bihonegn S., Agazie T., Marie A., Shiferaw S., Fentaw L., Mohammed A. Cosmetic textiles: important active ingredients, products and their applications (2025) Research Journal of Textile and Apparel. DOI: 10.1108/RJTA-08-2024-0140

46. Bezerra F.M., Zurita M.E.P.P., Volante E.K.T.S., Moisés M.P., Lis M.J. Microcapsules and Biofunctionality: Enhancing Textile and Dermocosmetic Properties Through Microencapsulation (2025) Journal of Applied Polymer Science. DOI: 10.1002/app.57036

47. Tariq H., Rehman A., Raza Z.A., Kishwar F., Abid S. Recent progress in the microencapsulation of essential oils for sustainable functional textiles (2024) Polymer Bulletin, 81 (9), pp. 7585 – 7629. DOI: 10.1007/s00289-023-05092-x

48. Singh M.K., Singh A., Morris H.V. Cosmeto-textiles (2023) Textile Progress, 55 (3), pp. 109 – 163. DOI: 10.1080/00405167.2023.2258675

49. Julaeha E., Nurzaman M., Wahyudi T., Nurjanah S., Permadi N., Anshori J.A. The Development of the Antibacterial Microcapsules of Citrus Essential Oil for the Cosmetotextile Application: A Review (2022) Molecules, 27 (22), art. no. 8090. DOI: 10.3390/molecules27228090

50. Ben Abdelkader, M., Azizi, N., Baffoun, A., Chevalier, Y., Majdoub, M. New microcapsules based on isosorbide for cosmetotextile: Preparation and characterization (2018) Industrial Crops and Products, 123, pp. 591-599. DOI: 10.1016/j.indcrop.2018.07.020

51. Kou, S., Chang, S., Huang, Z., Min, X., Zhang, X., Fang, M., Hao, L. Flexible Phase Change Composite Fibers for Wearable Applications—A Review (2025) Journal of Applied Polymer Science. DOI: 10.1002/app.57962

52. Peng, X., Umer, M., Pervez, M.N., Hasan, K.M.F., Habib, M.A., Islam, M.S., Lin, L., Xiong, X., Naddeo, V., Cai, Y. Biopolymers-based microencapsulation technology for sustainable textiles development: A short review (2023) Case Studies in Chemical and Environmental Engineering, 7, art. no. 100349. DOI: 10.1016/j.cscee.2023.100349

53. Ghonia J.R., Savani N.G., Prajapati V., Dholakiya B.Z. A review on polyurethane based multifunctional materials synthesis for advancement in textile coating applications (2024) Journal of Polymer Research, 31 (3), art. no. 95. DOI: 10.1007/s10965-024-03941-5

54. Yılmaz, B., Karavana, H.A. Application of chitosan-encapsulated orange oil onto footwear in sock leathers spray drying technique for an environmentally sustainable antibacterial formulation (2020) Johnson Matthey Technology Review, 64 (4), pp. 443 – 451. DOI: 10.1595/205651320X15901340190139

55. Silva, M., Martins, I.M., Barreiro, M.F., Dias, M.M., Egídio Rodrigues, A.E. Functionalized textiles with PUU/limonene microcapsules: effect of finishing methods on fragrance release (2017) Journal of the Textile Institute, 108 (3), pp. 361 – 367. DOI: 10.1080/00405000.2016.1166823

56. Zhang, Y., Zhang, W., Wang, X., Dong, Q., Zeng, X., Quirino, R.L., Lu, Q., Wang, Q., Zhang, C. Waterborne polyurethanes from castor oil-based polyols for next generation of environmentally-friendly hair-styling agents (2020b) Progress in Organic Coatings, 142, art. no. 105588,. DOI: 10.1016/j.porgcoat.2020.105588

57. Zareanshahraki, F., Mannari, V. “Green” UV-LED gel nail polishes from bio-based materials (2018) International Journal of Cosmetic Science, 40 (6), pp. 555-564. DOI: 10.1111/ics.12497

58. Hedayati S., Tarahi M., Iraji A., Hashempur M.H. Recent developments in the encapsulation of lavender essential oil (2024) Advances in Colloid and Interface Science, 331, art. no. 103229. DOI: 10.1016/j.cis.2024.103229

59. Lobel B.T., Baiocco D., Al-Sharabi M., Routh A.F., Zhang Z., Cayre O.J. Current Challenges in Microcapsule Designs and Microencapsulation Processes: A Review (2024) ACS Applied Materials and Interfaces, 16 (31), pp. 40326 – 40355. DOI: 10.1021/acsami.4c02462

60. Giustra M., Sinesi G., Spena F., De Santes B., Morelli L., Barbieri L., Garbujo S., Galli P., Prosperi D., Colombo M. Microplastics in Cosmetics: Open Questions and Sustainable Opportunities (2024) ChemSusChem, 17 (22), art. no. e202401065. DOI: 10.1002/cssc.202401065

61. Elkalla E., Khizar S., Tarhini M., Lebaz N., Zine N., Jaffrezic-Renault N., Errachid A., Elaissari A. Core-shell micro/nanocapsules: from encapsulation to applications (2023) Journal of Microencapsulation, 40 (3), pp. 125 – 156. DOI: 10.1080/02652048.2023.2178538

62. Kłosowska A., Wawrzyńczak A., Feliczak-Guzik A. Microencapsulation as a Route for Obtaining Encapsulated Flavors and Fragrances (2023) Cosmetics, 10 (1), art. no. 26. DOI: 10.3390/cosmetics10010026

63. Li J., Parakhonskiy B.V., Skirtach A.G. A decade of developing applications exploiting the properties of polyelectrolyte multilayer capsules (2023) Chemical Communications, 59 (7), pp. 807 – 835. DOI: 10.1039/d2cc04806j

64. Russell S., Bruns N. Encapsulation of Fragrances in Micro- and Nano-Capsules, Polymeric Micelles, and Polymersomes (2023) Macromolecular Rapid Communications, 44 (16), art. no. 2300120. DOI: 10.1002/marc.202300120

65. Parente J.F., Sousa V.I., Marques J.F., Forte M.A., Tavares C.J. Biodegradable Polymers for Microencapsulation Systems (2022) Advances in Polymer Technology, 2022, art. no. 4640379. DOI: 10.1155/2022/4640379

66. Anıl, D., Berksun, E., Durmuş-Sayar, A., Sevinis, E. B., Ünal, S. Recent advances in waterborne polyurethanes and their nanoparticle-containing dispersions (2020). *Handbook of waterborne coatings*, 249-302. DOI: 10.1016/B978-0-12-814201-1.00011-1

67. Qiu J., Zhao H., Luan S., Wang L., Shi H. Recent advances in functional polyurethane elastomers: from structural design to biomedical applications (2025) Biomaterials Science. DOI: 10.1039/d5bm00122f

68. Zainab, I., Naseem, Z., Batool, S.R., Waqas, M., Nazir, A., Nazeer, M.A. Polyurethane/silk fibroin-based electrospun membranes for wound healing and skin substitute applications (2025) Beilstein Journal of Nanotechnology, 16, pp. 591–612. DOI:10.3762/bjnano.16.46

69. Hebda E., Pielichowski K. Biomimetic Polyurethanes in Tissue Engineering (2025) Biomimetics, 10 (3), art. no. 184. DOI: 10.3390/biomimetics10030184

70. Shanno K., Mangala P., Shanmugarajan T.S., Bhyan B., Shinde M.G., Rane B.Y., Ali S.S., Kumar M., Kumar P. 3D-Printed Polyurethane Scaffolds for Bone Tissue Engineering: Techniques and Emerging Applications (2025) Regenerative Engineering and Translational Medicine. DOI: 10.1007/s40883-024-00381-x

71. Joseph T.M., Thomas M.G., Mahapatra D.K., Unni A.B., Kianfar E., Haponiuk J.T., Thomas S. Adaptive and intelligent polyurethane shape-memory polymers enabling next-generation biomedical platforms (2025) Case Studies in Chemical and Environmental Engineering, 11, art. no. 101165. DOI: 10.1016/j.cscee.2025.101165

72. Azarmgin S., Torabinejad B., Kalantarzadeh R., Garcia H., Velazquez C.A., Lopez G., Vazquez M., Rosales G., Heidari B.S., Davachi S.M. Polyurethanes and Their Biomedical Applications (2024) ACS Biomaterials Science and Engineering, 10 (11), pp. 6828 – 6859. DOI: 10.1021/acsbiomaterials.4c01352

73. Zhang J., Lv S., Zhao X., Ma S., Zhou F. Functional Zwitterionic Polyurethanes: State-of-the-Art Review (2024) Macromolecular Rapid Communications, 45 (5), art. no. 2300606. DOI: 10.1002/marc.202300606

74. Wendels, S., Avérous, L. Biobased polyurethanes for biomedical applications (2021) Bioactive Materials, 6 (4), pp. 1083-1106. DOI: 10.1016/j.bioactmat.2020.10.002

75. Sikdar, P., Dip, T.M., Dhar, A.K., Bhattacharjee, M., Hoque, M.S., Ali, S.B. Polyurethane (PU) based multifunctional materials: Emerging paradigm for functional textiles, smart, and biomedical applications (2022) Journal of Applied Polymer Science, e52832. DOI: 10.1002/app.52832

76. Feldman, D. Polyurethane and polyurethane nanocomposites: Recent contributions to medicine (2021) Biointerface Research in Applied Chemistry, 11 (1), pp. 8179-8189. DOI: 10.33263/BRIAC111.81798189

77. Gupta, A., Maharjan, A., Kim, B.S. Shape memory polyurethane and its composites for various applications (2019) Applied Sciences (Switzerland), 9 (21), art. no. 4694,. DOI: 10.3390/app9214694

78. Balavigneswaran, C.K., Muthuvijayan, V. Nanohybrid-Reinforced Gelatin-Ureidopyrimidinone-Based Self-healing Injectable Hydrogels for Tissue Engineering Applications (2021) ACS Applied Bio Materials, 4 (6), pp. 5362-5377. DOI: 10.1021/acsabm.1c00458

79. Mohamady Hussein, M.A., Gunduz, O., Sahin, A., Grinholc, M., El-Sherbiny, I.M., Megahed, M. Dual Spinneret Electrospun Polyurethane/PVA-Gelatin Nanofibrous Scaffolds Containing Cinnamon Essential Oil and Nanoceria for Chronic Diabetic Wound Healing: Preparation, Physicochemical Characterization and In-Vitro Evaluation (2022) Molecules, 27 (7), art. no. 2146,. DOI: 10.3390/molecules27072146

80. Ma, Y., Hu, N., Liu, J., Zhai, X., Wu, M., Hu, C., Li, L., Lai, Y., Pan, H., Lu, W.W., Zhang, X., Luo, Y., Ruan, C. Three-Dimensional Printing of Biodegradable Piperazine-Based Polyurethane-Urea Scaffolds with Enhanced Osteogenesis for Bone Regeneration (2019) ACS Applied Materials and Interfaces, 11 (9), pp. 9415-9424. DOI: 10.1021/acsami.8b20323

81. Mu, Y., Lv, S., Liu, J., Tong, J., Liu, L., Wang, J., He, T., Wei, D. Recent advances in research on biomass-based food packaging film materials (2025) Comprehensive Reviews in Food Science and Food Safety, 24, e70093. DOI:10.1111/1541-4337.70093

82. Song W., Muhammad S., Dang S., Ou X., Fang X., Zhang Y., Huang L., Guo B., Du X. The state-of-art polyurethane nanoparticles for drug delivery applications (2024) Frontiers in Chemistry, 12, art. no. 1378324. DOI: 10.3389/fchem.2024.1378324

83. Niu, Y., Stadler, F.J., Song, J., Chen, S., Chen, S. Facile fabrication of polyurethane microcapsules carriers for tracing cellular internalization and intracellular pH-triggered drug release (2017) Colloids and Surfaces B: Biointerfaces, 153, pp. 160-167. DOI: 10.1016/j.colsurfb.2017.02.018

84. Mudri, N.H., Abdullah, L.C., Aung, M.M., Biak, D.R.A., Tajau, R. Structural and rheological properties of nonedible vegetable oil-based resin (2021) Polymers, 13 (15), art. no. 2490,. DOI: 10.3390/polym13152490

85. Szpiłyk, M., Lubczak, R., Lubczak, J. The biodegradable cellulose-derived polyol and polyurethane foam (2021) Polymer Testing, 100, art. no. 107250,. DOI: 10.1016/j.polymertesting.2021.107250

86. Wei, K., Wang, X., Ma, B., Shi, W., Duan, S., Liu, F. Study on rheological properties and phase-change temperature control of asphalt modified by polyurethane solid–solid phase change material (2019) Solar Energy, 194, pp. 893-902. DOI: 10.1016/j.solener.2019.11.007

87. Ali, M., Meaney, S.P., Abedin, M.J., Holt, P., Majumder, M., Tabor, R.F. Graphene oxide–silica hybrid capsules for sustained fragrance release (2019) Journal of Colloid and Interface Science, 552, pp. 528-539. DOI: 10.1016/j.jcis.2019.05.061

88. Xiao Z., Zhou L., Sun P., Li Z., Kang Y., Guo M., Niu Y., Zhao D. Regulation of mechanical properties of microcapsules and their applications (2024) Journal of Controlled Release, 375, pp. 90 – 104. DOI: 10.1016/j.jconrel.2024.09.001

89. Mallakpour, S., Tabesh, F., Hussain, C.M. 3D and 4D printing: From innovation to evolution (2021) Advances in Colloid and Interface Science, 294, art. no. 102482. DOI: 10.1016/j.cis.2021.102482

90. Shinde, S., Mane, R., Vardikar, A., Dhumal, A., Rajput, A. 4D printing: From emergence to innovation over 3D printing (2023) European Polymer Journal, 197, art. no. 112356. DOI: 10.1016/j.eurpolymj.2023.112356

91. Kamkar, M., Salehiyan, R., Goudoulas, T.B., Abbasi, M., Saengow, C., Erfanian, E., Sadeghi, S., Natale, G., Rogers, S.A., Giacomin, A.J., Sundararaj, U. Large amplitude oscillatory shear flow: Microstructural assessment of polymeric systems (2022) Progress in Polymer Science, 132, art. no. 101580,. DOI: 10.1016/j.progpolymsci.2022.101580

92. Kang C.-H., Guo J.-X., Fei D.-T., Kiyingi W. Intelligent responsive self-assembled micro-nanocapsules: Used to delay gel gelation time (2024) Petroleum Science, 21 (4), pp. 2433 – 2443. DOI: 10.1016/j.petsci.2024.04.011

93. Konieczny, J., Loos, K. Polyurethane Coatings Based on Renewable White Dextrins and Isocyanate Trimers (2019a) Macromolecular Rapid Communications, 40 (10), art. no. 1800874,. DOI: 10.1002/marc.201800874

94. Konieczny, J., Loos, K. Green polyurethanes from renewable isocyanates and biobased white dextrins (2019b) Polymers, 11 (2), art. no. 256,. DOI: 10.3390/polym11020256

95. Gillece, T., Mcmullen, R.L., Fares, H., Senak, L., Ozkan, S., Foltis, L. Probing the textures of composite skin care formulations using large amplitude oscillatory shear (2016) Journal of Cosmetic Science, 67 (3), pp. 121-159.

96. Dakhil H., Wierschem A. Measuring low viscosities and high shear rates with a rotational rheometer in a thin-gap parallel-disk configuration (2014) Applied Rheology, 24 (6), art. no. 63795 DOI: 10.3933/APPLRHEOL-24-63795

97. Dakhil H., Auhl D., Wierschem A. Infinite-shear viscosity plateau of salt-free aqueous xanthan solutions (2019) Journal of Rheology, 63 (1), pp. 63 – 69. DOI: 10.1122/1.5044732

98. Dakhil, H., Basu, S.K., Steiner, S., Gerlach, Y., Soller, A., Pan, S., Germann, N., Leidenberger, M., Kappes, B., Wierschem, A. Buffered λ-DNA solutions at high shear rates (2021) Journal of Rheology, 65 (2), pp. 159-169. DOI: 10.1122/8.0000136

99. Büyükurgancı B., Basu S.K., Neuner M., Guck J., Wierschem A., Reichel F. Shear rheology of methyl cellulose based solutions for cell mechanical measurements at high shear rates (2023) Soft Matter, 19 (9), pp. 1739 – 1748. DOI: 10.1039/d2sm01515c

100. Belwal, T., Chemat, F., Venskutonis, P.R., Cravotto, G., Jaiswal, D.K., Bhatt, I.D., Devkota, H.P., Luo, Z. Recent advances in scaling-up of non-conventional extraction techniques: Learning from successes and failures (2020) TrAC - Trends in Analytical Chemistry, 127, art. no. 115895,. DOI: 10.1016/j.trac.2020.115895

101. European Commission. Cosmetics. Accessed on 15 October, 2025. URL: https://single-market-economy.ec.europa.eu/sectors/cosmetics_en